\documentclass[prxquantum, reprint,superscriptaddress]{revtex4-2}
\usepackage{amsmath,amssymb,amsthm}
\usepackage[framemethod=tikz]{mdframed}
\usepackage{array}
\newcolumntype{C}[1]{>{\centering\arraybackslash}m{#1}}
\usepackage{natbib}
\usepackage{mathtools}
\usepackage{braket}
\usepackage[hidelinks,colorlinks=false]{hyperref}
\usepackage{graphicx}
\usepackage{color, xcolor}
\usepackage{bm}
\usepackage{makecell}
\DeclareMathOperator{\Tr}{Tr}

\usepackage[framemethod=tikz]{mdframed} 
\definecolor{lightgray}{gray}{0.90}
\newmdtheoremenv[
  style=definition,        
  linewidth=0.5pt,
  roundcorner=4pt,
  backgroundcolor=lightgray,
skipabove=6pt,skipbelow=6pt, innerleftmargin=6pt,innerrightmargin=6pt,
innertopmargin=-2pt,innerbottommargin=4pt,
]{boxedresultdef}{Result}
\begin{document}
\title{Symmetries of Pauli Noise from Lindbladian Dynamics}
\date{\today}

\author{Moein Malekakhlagh}
\email{moein.malekakhlagh@ibm.com}
\affiliation{IBM Quantum, IBM T.~J.~Watson Research Center, Yorktown Heights, NY 10598, USA}
\author{Edward H.~Chen}
\altaffiliation{Present address: Microsoft Discovery and Quantum, One Microsoft Way, Redmond, 98052, WA, USA}
\affiliation{IBM Quantum, IBM T.~J.~Watson Research Center, Yorktown Heights, NY 10598, USA}
\author{Luke C.~G.~Govia}
\altaffiliation{Present address: CMC Microsystems, Waterloo, ON, Canada.}
\affiliation{IBM Quantum, IBM T.~J.~Watson Research Center, Yorktown Heights, NY 10598, USA}
\author{Alireza Seif}
\email{alireza.seif@ibm.com}
\affiliation{IBM Quantum, IBM T.~J.~Watson Research Center, Yorktown Heights, NY 10598, USA}

\begin{abstract}
Characterizing noise in quantum circuits is fundamentally limited by gauge degrees of freedom; certain parameters, such as the individual contributions of state preparation and measurement (SPAM) errors, are in principle unlearnable from any experiment within the gate set. Here, we show that the physical structure of realistic noise processes imposes approximate symmetry constraints on the Pauli fidelities of gate noise channels. These symmetries relate the fidelity of a Pauli $P$ and its gate-conjugate $U_g P U_g ^{\dagger}$, and can be used to fix the gauge using only knowledge of the error type and not its magnitude. Using Lindbladian perturbation theory, we analyze a broad class of Clifford gates, including $ZZ_{\pi/2}$, CZ, CNOT, iSWAP, and SWAP, and demonstrate that coherent errors do not induce first-order asymmetry, while only a restricted set of predominantly off-diagonal dissipative errors can break the symmetry at first order, for which we derive simple selection rules. Notably, common single-qubit noise sources such as $T_1$-relaxation and $T_{2\phi}$-pure-dephasing can only cause asymmetry at second order. Leveraging these symmetries to fix the gauge enables systematic identification of SPAM errors, simplifying error characterization and mitigation. We validate our results numerically and experimentally on IBM Kingston.
\end{abstract}
\maketitle

\section{Introduction}
\label{Sec:Intro}
Accurate models of hardware noise underpin a wide range of tasks across quantum computing, including error mitigation, quantum error correction, and hardware characterization. Many of these applications rely on the noise being well described by a Pauli channel. Randomized compiling and Pauli twirling make this the case, tailoring the effective noise on Clifford gates to Pauli channels~\cite{geller2013efficient, wallman2016noise, hashim2021randomized}, which can then be learned at scale using cycle benchmarking~\cite{erhard2019characterizing, flammia2020efficient} or sparse Pauli--Lindblad (SPL) models~\cite{vandenberg2023probabilistic, berg2023techniques}. In error mitigation~\cite{temme2017error, cai2023quantum}, these learned channels feed directly into probabilistic error cancellation (PEC)~\cite{temme2017error, vandenberg2023probabilistic} and zero-noise extrapolation (ZNE)~\cite{kim2023evidence}. In quantum error correction, noise-aware decoders exploit the learned structure to improve logical error rates~\cite{hockings2025improving}. Additionally, noise learning is also central to hardware characterization, where the goal is not only to predict circuit outcomes but also to identify the magnitude and structure of individual error mechanisms. Such physically interpretable estimates are essential for diagnosing devices and guiding hardware improvement.
% Most quantum error mitigation techniques rely on accurate models of hardware noise~\cite{temme2017error, cai2023quantum}. Randomized compiling and Pauli twirling tailor the effective noise on Clifford gates to Pauli channels~\cite{geller2013efficient, wallman2016noise, hashim2021randomized},
% which can then be learned at scale using cycle benchmarking~\cite{erhard2019characterizing, flammia2020efficient} or sparse Pauli--Lindblad (SPL)
% models~\cite{vandenberg2023probabilistic, berg2023techniques}. These learned channels feed directly into probabilistic error cancellation (PEC)~\cite{temme2017error, vandenberg2023probabilistic} and zero-noise extrapolation (ZNE)~\cite{kim2023evidence}. Beyond mitigation, noise learning is also central to hardware characterization, where the goal is not only to predict circuit outcomes but also to identify the magnitude and structure of individual error mechanisms. Such physically interpretable estimates are essential for diagnosing devices and guiding hardware improvement.

However, not all parameters of a Pauli noise model can be determined from experiment. It is well established that gauge degrees of freedom prevent unique identification of the noise affecting state preparation (SP), measurement (M), and gates~\cite{nielsen2021gate, chen2023learnability,
chen2026efficient}. Concretely, a generalized depolarizing channel and its inverse can be freely inserted between noise channels without changing any observable outcome. For $n$ qubits under a quasi-local noise ansatz, $n$ single-qubit depolarizing parameters remain undetermined~\cite{chen2026efficient}.

In current practice, the gauge is typically fixed by assuming that conjugate Pauli fidelities of the gate channel are symmetric, and by treating state preparation as noiseless when calibrating measurement
errors~\cite{vandenberg2023probabilistic, vandenberg2022model}. These assumptions fix each component of the gate set independently, which can produce inconsistent noise models. Recent theoretical and experimental work has shown that such inconsistencies lead to biased error-mitigated expectation values~\cite{chen2026efficient, chen2025disambiguating}. The self-consistent framework of Refs.~\cite{chen2026efficient, chen2025disambiguating} resolves this by learning all noise channels under a shared gauge, yielding provably unbiased PEC estimators for any gauge choice. That approach, however, leaves the gauge as a free parameter.

Other approaches to resolving the SPAM gauge ambiguity have been proposed. Ancilla-based schemes can separately quantify state-preparation and measurement error rates by correlating a target qubit with a neighboring qubit, assuming that entangling gate errors can be independently mitigated~\cite{yu2025efficient}. On platforms with accessible higher energy levels, qutrit or qudit control can be used to tighten positivity constraints and reduce the gauge ambiguity~\cite{chen2025enhancing, haupt2025mitigating}. These methods rely on additional hardware resources, either ancilla qubits or noncomputational energy levels, and make assumptions about the quality and noise of the associated control
operations.

%%%%%%%%%%%%%%% Fig: schematic %%%%%%%%%%%%%
\begin{figure*}
\centering
\includegraphics[scale=0.160]{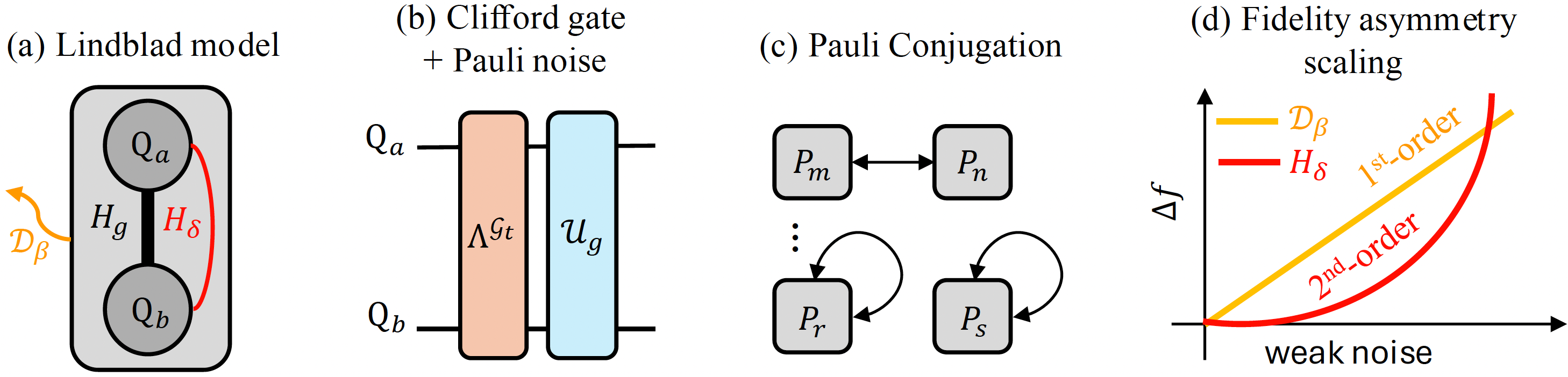}
\caption{\textbf{Physical origin and characterization of Pauli noise symmetry}---(a) Open-system Lindblad description of a two-qubit gate with ideal Hamiltonian $H_g$, noise Hamiltonian $H_{\delta}$, and noise dissipator $\mathcal{D}_{\beta}$. (b) The corresponding Clifford gate $\mathcal{U}_g$ and (\textit{twirled}) Pauli noise channel $\Lambda^{\mathcal{G}_{\text{t}}}$. We analyze the resulting Pauli noise channel both numerically (Appendix~\ref{App:AsymmSim}) and analytically via Lindblad perturbation theory following Ref.~\cite{malekakhlagh2025efficient} (Appendices~\ref{App:LindPert}--\ref{App:MultiPauliWithPhase2}). (c) Under the action of a Clifford gate, Pauli operators are permuted according to $U_g P_n U_g^{\dagger} = P_m$, typically forming cycles of length 1 or 2. A symmetry measure of interest is the difference in the channel fidelity of the paired Paulis as $\Delta f_{mn} = f_m - f_n$. (d) We find that the fidelity asymmetry $\Delta f$ depends only weakly on Hamiltonian noise parameters $\boldsymbol{\delta}$ at second order, whereas certain dissipative noise components—predominantly off-diagonal elements in the Pauli basis—can induce a first-order asymmetry in $\boldsymbol{\beta}$. We present effective selection rules for this purpose based on the structure of the physical gate generator (Results~\ref{result:OnePauli}--\ref{result:MultiPauliPhase2} and Fig.~\ref{fig:dissipator_asymmetry_patterns}).}
\label{fig:schematic}
\end{figure*}
%%%%%%%%%%%%%%%%%%%%%%%%%%%%%%%%%%%%%%%%%%%%%

In this work, we take a different approach that requires no additional hardware resources and instead exploits the physical structure of the noise to fix the gauge. We analyze the Pauli-fidelity
symmetry of two-qubit Clifford gates, i.e., whether the fidelity of a Pauli operator and that of its transformation under the gate are equal, using Lindblad
perturbation theory applied to a general noise model with both coherent and dissipative contributions \cite{Gorini_completely_1976, Lindblad_Generators_1976, Breuer_Theory_2002, Blume_Taxonomy_2022, malekakhlagh2025efficient, miller2025efficient, berg2025large} (Fig.~\ref{fig:schematic}). This continuous-time description is essential: rather than attaching a fixed noise channel before or after the ideal operation, it lets the noise evolve throughout the gate in the interaction frame of the ideal Hamiltonian, which is precisely what determines the symmetry and asymmetry selection rules we derive here. We find that for several widely used two-qubit gates such as $ZZ_{\pi/2}$, CZ, CNOT, iSWAP, and SWAP, Hamiltonian errors produce fidelity asymmetry only at second order in the noise strength, while first-order asymmetry arises exclusively from certain, mainly off-diagonal, dissipator elements in the Pauli basis satisfying selection rules tied to the gate generator. Standard single-qubit $T_1$-relaxation and $T_{2\phi}$-dephasing dissipators produce only second-order asymmetry.

Together, these results motivate a physically informed gauge-fixing procedure based on imposing a first-order fidelity symmetry on the gate channel. This prescription requires only knowledge of the error type, not its magnitude, and thereby provides a practical route to resolving the gauge ambiguity in SPAM characterization. Once the gauge is fixed, state-preparation and measurement errors become separately identifiable and can be mitigated independently. This separation avoids the bias identified in Ref.~\cite{chen2025disambiguating}, which can arise when probabilistic error cancellation (PEC) methods that rely on symmetry assumptions are combined with measurement-error-mitigation protocols, such as Twirled Readout Error eXtinction (TREX)~\cite{vandenberg2022model}, that assume perfect state preparation. We experimentally demonstrate this approach on IBM Kingston, where the physically motivated gauge separates SPAM errors into distinct state-preparation and measurement contributions. The extracted error budgets reveal a consistent hierarchy in which state-preparation errors are smaller than measurement errors, in some cases by nearly an order of magnitude, in agreement with the expected physics of superconducting qubit devices.

Beyond SPAM characterization, these results also have direct implications for circuit-level noise simulation. Standard simulations often insert noise channels either before or after an ideal gate, even though physical noise acts continuously during gate operation \cite{malekakhlagh2025efficient}. Our results show that Clifford gates effectively symmetrize common physically motivated noise processes during the gate. As a result, microscopic error mechanisms such as amplitude damping or dephasing should first be mapped to their gate-symmetrized effective channels before being used in a circuit model. These effective channels can then be inserted before or after the ideal gate, providing a more realistic simulation primitive that captures the leading effect of in-gate noise without requiring explicit integration of the microscopic dynamical generator.

%%%%%%%%%Sec: Lindblad noise model %%%%%%%%%%%%%%%%%%
\section{Continuous-time Lindblad noise model}
\label{Sec:LindModel}
We describe noisy gate evolution \cite{ Blume_Taxonomy_2022, malekakhlagh2025efficient, miller2025efficient, berg2025large} using the Lindblad master equation \cite{Gorini_completely_1976, Lindblad_Generators_1976, Breuer_Theory_2002}
\begin{align}
\dot{\rho}(t) = \mathcal{L}\rho(t)
= -i[H_g + H_{\delta},\,\rho(t)] + \mathcal{D}_{\beta}\rho(t),
\label{eq:LindEq}
\end{align}
where $H_g$ denotes the ideal gate Hamiltonian and 
$H_{\delta} \equiv \tfrac{1}{2}\sum_{j} \delta_{j} P_{j}$ represents coherent noise generated by Pauli operators $P_j$ with real amplitudes $\delta_j$. Dissipative errors are parametrized generically as
\begin{align}
\mathcal{D}_{\beta}\,\rho(t)\equiv 
\sum_{jk} \beta_{jk}\!\left(
P_j \rho(t) P_k^{\dagger}
- \tfrac{1}{2}\{ P_k^{\dagger} P_j, \rho(t) \}
\right),
\label{eq:Def_of_Dbeta}
\end{align}
with $\bm{\beta}$ specifying the dissipative noise rates (Fig.~\ref{fig:schematic}(a)). Markovianity requires $\bm{\beta}$ to be positive semi-definite (PSD).  

Using this framework, we analyze the resulting noise channels and the symmetry structure of Pauli fidelities for Clifford two-qubit gates. The corresponding gate noise channel is obtained as
\begin{align}
\Lambda^{\mathcal{G}}\equiv\mathcal{U}_{g}^{-1}\mathcal{G} \;,
\label{eq:Def_of_Lambda^G}
\end{align} 
where the ideal and noisy evolution over the gate duration $\tau_g$ are defined respectively as 
\begin{align}
&\mathcal{U}_{g} = e^{-i\mathcal{H}_{g}\tau_{g}} \;, \\
&\mathcal{G} = e^{\mathcal{L}\tau_{g}} \;,
\end{align}
with $\mathcal{H}$ denoting the superoperator representation of $H$. For Clifford gates, the noise channel can be twirled \cite{geller2013efficient, wallman2016noise,hashim2021randomized} into an effective Pauli noise channel by conjugating with Pauli gates:
\begin{align}
\Lambda^{\mathcal{G}_t}\equiv \mathcal{T}_{P}(\Lambda^{\mathcal{G}}) = \frac{1}{|\mathcal{P}_N|}\sum\limits_{\mathcal{P}\in \mathcal{P}_N} \mathcal{P}^{\dag}\Lambda^{\mathcal{G}}\mathcal{P} \;.
\label{eq:Def_of_Lambda^Gt}
\end{align}
In the Pauli Transfer Matrix (PTM) representation \cite{gambetta2012characterization, greenbaum2015introduction}, this corresponds to retaining only the diagonal entries of the channel. We define the corresponding fidelity of a Pauli operator $P_i$ as
\begin{align}
f_i \equiv \frac{1}{D}\Tr\{P_i \Lambda^{\mathcal{G}_{\text{t}}}(P_i)\},
\label{eq:Def_of_f}
\end{align}
for Hilbert-space dimension $D=2^{N}$ and qubit number $N=2$.

Our analysis focuses on the leading-order dependence of the effective Pauli noise channel parameters on the underlying physical noise strengths $\boldsymbol{\delta}$ and $\boldsymbol{\beta}$. These dependencies are derived analytically using a Lindblad-Dyson perturbative expansion~\cite{malekakhlagh2025efficient} (Appendix~\ref{App:LindPert}) and are further validated through numerical simulations (Appendix~\ref{App:AsymmSim}). To leading order, the noise channel associated with gate \(\mathcal{G}\) is approximated as
\begin{align}
\Lambda^{\mathcal{G}} \equiv e^{\Omega(\tau_g,0)} \approx \mathcal{I} + \Omega_1(\tau_g,0) + O(\mathcal{L}_I^2) \;,  
\end{align}
where the first-order generator is given by
\begin{align}
\Omega_1(\tau_g,0) \equiv \int_{0}^{\tau_g}dt' \mathcal{L}_I(t') \;.
\end{align}
Here, $\mathcal{L}_I(t)$ denotes the noise Lindbladian expressed in the interaction frame of the ideal gate (Appendix~\ref{App:LindPert}). We then extract the diagonal elements of the noise channel PTM to characterize the symmetries of the resulting Pauli channel $\Lambda^{\mathcal{G}_t}$ (Fig.~\ref{fig:schematic}(b)).

In the following, we apply this framework to study representative families of Clifford gates and derive selection rules that determine when dissipative noise gives rise to first-order fidelity asymmetries.  
%%%%%%%%%%%%%%%%%%%%%%%%%%  

%%%%%%%%%%%%% Sec:Symmetries of Pauli noise %%%%%%%%%%%%%%%%%%%%
\section{Fidelity symmetry of Clifford two-qubit gates}
\label{Sec:PauliSymmetry}
Pauli twirling is routinely used to map the noise of Clifford gates to an effective Pauli channel, which can then be learned using protocols such as cycle benchmarking ~\cite{erhard2019characterizing, flammia2020efficient}. Because Clifford gates permute Pauli operators under conjugation, Pauli fidelities are organized into permutation cycles of this action; only cycle-invariant combinations (e.g., products of fidelities along each cycle) are learnable \cite{chen2023learnability}, rather than the individual fidelities. For many standard two-qubit Clifford gates these orbits are predominantly 2-cycles (up to Pauli phases), in which case fidelities appear in fixed pairs and it is common to assume pairwise symmetry (Fig.~\ref{fig:schematic}(c)) \cite{vandenberg2023probabilistic}. 
Using a physically motivated Lindblad noise model, we assess the validity and limits of this assumption across representative entangling Clifford gates. Common elementary two-qubit Clifford gates admit a (non‑unique) representation as the exponential of a sum of mutually commuting Pauli operators \cite{gottesman1997heisenberg, dehaene2003clifford, aaronson2004improved}. 

\begin{figure*}[t!]
\centering
\includegraphics[scale=0.900]{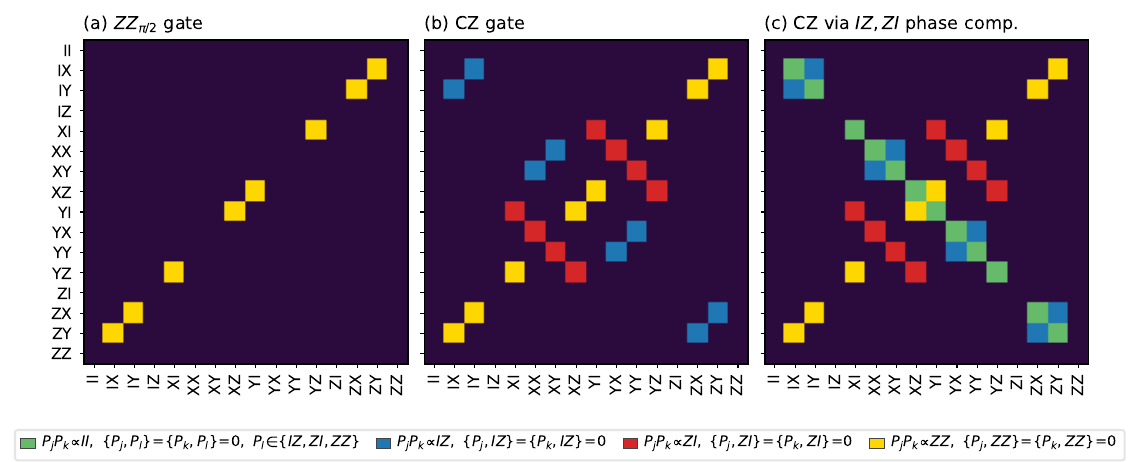}
\caption{\textbf{Leading-order pattern of Pauli-fidelity asymmetry induced by dissipative noise}---Shown are the dissipator entries $\boldsymbol{\beta}$ that produce a nonzero first-order Pauli-fidelity asymmetry for (a) $ZZ_{\pi/2}$, (b) CZ, and (c) CZ gate with single-qubit $IZ$ and $ZI$ phase compensation. For gates whose generators decompose into commuting Pauli terms, contributions are grouped by generator component; the patterns follow our selection rules in Results~\ref{result:OnePauli}, \ref{result:MultiPauli}, and \ref{result:MultiPauliPhase2}, respectively.}
\label{fig:dissipator_asymmetry_patterns}
\end{figure*}

We organize our analysis into families of Clifford gates according to the number of Pauli generators involved, and whether the gate is generated by a physically implemented common‑angle Hamiltonian or instead requires post‑gate correction, as: (1) single-Pauli generator gates such as $ZZ_{\pi/2}$; (2) gates generated by sums of commuting Paulis. The latter can be subdivided into three categories: 
(2a) commuting terms driven by a common rotation angle, as in idealized CZ, CNOT, iSWAP, and SWAP gates; and (2b) cases with unbalanced physical interactions where post-gate (single-qubit) corrections transform all terms into a common-angle gate such as a CZ gate with residual $IZ$ and $ZI$ terms as implemented in our experiments. The symmetry rules for these scenarios are summarized in Results~\ref{result:OnePauli}--\ref{result:MultiPauliPhase2} and visualized in Fig.~\ref{fig:dissipator_asymmetry_patterns}. We also analyze the complementary case (2c) in which a single Pauli term generates the target gate, while the remaining commuting terms appear as large but correctable coherent contributions, such as a $ ZZ_{\pi/2}$ gate with residual $IZ$ and $ZI$ terms (Result~\ref{result:MultiPauliPhase1} in Appendix~\ref{App:MultiPauliWithPhase1}). Cases (2b)--(2c) are motivated by the fact that physical couplings typically generate additional terms alongside the desired two-qubit interaction, rather than just an isolated entangling term \cite{Yan_Tunable_2018, Stehlik_Tunable_2021, Foxen_Demonstrating_2020, Sung_Realization_2021}. While not exhaustive over the two-qubit Clifford group, our categorization captures the standard elementary two-qubit gate sets used in practice.

%%%%%%%% Fig: dissipator asymmetry patern %%%%%%%%%%%%%
%%%%%%%%%%%%%%%%%%%%%%%%%%%%%%%%%%%%%%%%

Through perturbative methods and numerical simulations we show that coherent (Hamiltonian) noise yields Pauli fidelity asymmetries only at $O(\delta^{2})$ (Appendix~\ref{App:OnePauliGates}), whereas specific dissipator elements can induce first-order asymmetries at $O(\beta)$, depending on their commutation relations with the gate generator (Appendices~\ref{App:OnePauliGates}--\ref{App:MultiPauliWithPhase2}). From perturbative analysis, we derive simple selection rules identifying which dissipator terms produce such leading-order asymmetry (Fig.~\ref{fig:schematic}(d)). Notably, standard single-qubit $T_{1}$-relaxation and $T_{2\phi}$-dephasing channels contribute only at second order (Appendix~\ref{App:AsymmSim}).    

In particular, suppose that two Pauli operators $P_m$ and $P_n$ are mapped to one another under conjugation by the ideal gate as $P_m \propto U_g^\dagger P_n U_g$. For the Clifford gates considered here, which are generated by mutually commuting Pauli operators, this conjugation can be expressed as $P_m \propto U_g^{\dag}P_nU_g \propto P_l P_n$, where $P_l$ is either one generator of $U_g$ or a product of multiple generators, depending on which generators anticommute with $P_n$ (Appendices~\ref{App:OnePauliGates}--\ref{App:MultiPauliWithPhase2}). Using the Lindblad-Dyson expansion, we show that the first-order contribution to the Pauli fidelity asymmetry $\Delta f_{mn} = f_m - f_n$ can be expressed as (Appendix~\ref{App:OnePauliGates})
\begin{align}
\begin{split}
\Delta f^{(1)}_{mn,jk} =\frac{2}{D}\!\int_0^{\tau_g} dt' \beta_{jk} 
\Tr \Big[P_m^\dagger \Pi_{P_l}^- \big(P_j(t')\big) P_m\,P_{k}^\dagger(t')\Big].
\label{eq:Deltaf-compact}
\end{split}
\end{align}
Here, $\Delta f^{(1)}_{mn,jk}$ denotes the first-order fidelity asymmetry due to the dissipator element $\beta_{jk}$, $\Pi_{P_l}^-(\bullet)\equiv\tfrac{1}{2}(\bullet-P_l^\dagger \bullet P_l)$ is the projector onto the \emph{odd} part under $P_l$, and $P_j(t')$ and $P_k(t')$ are the corresponding interaction-frame Paulis evolved in the frame of the ideal $H_g$. In the following, we apply Eq.~\eqref{eq:Deltaf-compact} to the Clifford-gate categories introduced above.

\subsection{Gates generated by a single Pauli term} 
Consider a Clifford gate generated by a single Pauli operator as $H_g=(\omega_g/2)P_g$, where $\omega_g$ is the physical gate coupling strength. We show that the leading-order contribution to the fidelity asymmetry $\Delta f_{mn}$ for Pauli pairs $P_m \propto P_g P_n$ comes from off-diagonal dissipator elements that obey commutation relations $\{P_j,P_g\}=\{P_k,P_g\}=0$, as (Appendix~\ref{App:OnePauliGates})
\begin{align}
\Delta f_{mn,jk}^{(1)} = - 2 i \frac{\beta_{jk}}{D} \Tr\{P_gP_mP_jP_mP_k\}\int_{0}^{\tau_g} dt' \sin[2 \theta(t')],
\label{eq:Delf_mn offdiag_D Sol}
\end{align}
where $\theta(t')=\omega_g t'$, with the gate duration $\tau_g$ chosen such that $\omega_g \tau_g=\pi/2$. 

Expression~\eqref{eq:Delf_mn offdiag_D Sol} is non-zero only due to those $\bm{\beta}$ elements for which $P_jP_k \propto P_g$. We therefore identify the fidelity-asymmetry signature for this class of single-Pauli–generated Clifford gates as follows:  
\begin{boxedresultdef}[]
\label{result:OnePauli}
For a Clifford gate generated by the Hamiltonian $H_g = (\omega_g/2)\, P_g$, subject to arbitrary coherent and dissipative noise as in Eqs.~(\ref{eq:LindEq})--(\ref{eq:Def_of_Dbeta}), first-order Pauli-fidelity asymmetry arises solely from the off-diagonal dissipator elements $\beta_{jk}$ satisfying
\begin{align}
P_j P_k \propto P_g, \quad 
\{P_j, P_g\} = \{P_k, P_g\} = 0 \;.
\end{align}
\end{boxedresultdef}
For the example of a $ZZ_{\pi/2}$ gate, the first-order asymmetry comes from the dissipator elements $\beta_{jk}$ with $(j,k)$ and $(k,j)\in\{(IX,ZY), (IY,ZX), (XI,YZ), (YI,XZ)\}$, as shown in Fig.~\ref{fig:dissipator_asymmetry_patterns}(a).

\subsection{Gates generated by common-angle commuting Pauli terms}
In this setting, we show Result~\ref{result:OnePauli} extends naturally. The first‑order asymmetry decomposes into independent contributions from each commuting generator term (Appendix~\ref{App:MultiPauliGates}): 
\begin{boxedresultdef}[]
\label{result:MultiPauli}
For a Clifford gate generated by the Hamiltonian $H_g = \sum_{l\in G} (\omega_g/2)P_{l}$ where $[P_l,P_{l'}]=0$, $\forall l,l'\in G$, subject to arbitrary coherent and dissipative noise as in Eqs.~(\ref{eq:LindEq})--(\ref{eq:Def_of_Dbeta}), first-order Pauli-fidelity asymmetry arises from off-diagonal dissipator elements $\beta_{jk}$ satisfying
\begin{align}
P_j P_k \propto P_l, \quad 
\{P_j, P_l\} = \{P_k, P_l\} = 0 \;,
\end{align}
for any non-identity Pauli operators $P_l \in G$.
\end{boxedresultdef}
For the CZ gate $=\exp[-i(\pi/4)(II-IZ-ZI+ZZ)]$, the three commuting generator terms $\{IZ, ZI , ZZ\}$ define three independent eight-element subspaces, as shown in Fig.~\ref{fig:dissipator_asymmetry_patterns}(b). Other notable gates in this family include CNOT, iSWAP, and SWAP, each generated by a set of commuting Pauli terms. In each case, the commuting structure of the generators leads to distinct patterns in the asymmetric dissipator elements. For example, for the CNOT gate the leading-order contributions arise from selection rules set by the generator terms $\{ZI, IX , ZX\}$. Similarly, the asymmetry pattern of the iSWAP and SWAP gates are determined by the generator set $\{XX, YY\}$, and $\{XX, YY , ZZ\}$, respectively (Appendix~\ref{App:AsymmSim}).    

\subsection{Gates generated by multiple commuting Pauli terms corrected into a common-angle form}
\label{SubSec:MultiPauliWithPhase2}
We now consider a more general setting, where the \textit{physical} Hamiltonian contains multiple mutually commuting Pauli terms with unequal strength. This is the setting realized in our experiments in Sec.~\ref{Sec:Experiment} for CZ gates with residual $IZ$ and $ZI$ contributions. Here, we coherently \textit{retarget} the operation into a standard common-angle Clifford CZ gate by applying end-of-gate (single-qubit) Pauli corrections. Throughout this analysis, we assume that the correction gates are noiseless and applied outside the noisy gate evolution.

We summarize the asymmetry pattern for this case as follows (Appendix~\ref{App:MultiPauliWithPhase2}): 
\begin{boxedresultdef}
\label{result:MultiPauliPhase2}
Consider a Clifford gate which is \textit{physically} generated by $H_g = \sum_{l\in G} (\omega_g/2)\alpha_l P_{l}$ with $[P_l,P_{l'}]=0$ for any $l,l'$, and let $\mathcal{A}=\langle \{P_l\}_{l\in G}\rangle$ be the Abelian subgroup they generate (including identity). Assume the target Clifford gate is a common-angle rotation implemented via end-of-gate compensation $e^{i (\phi_l/2) P_l}$ with $\phi_l \equiv (\pm1-\alpha_l)\omega_g\tau_g$. For arbitrary coherent and dissipative noise as in Eqs.~(\ref{eq:LindEq})--(\ref{eq:Def_of_Dbeta}),
the first-order Pauli-fidelity asymmetry comes from dissipator elements $\beta_{jk}$ satisfying
\begin{align}
P_jP_k \propto P_l , \quad \{P_j,P_l\}=\{P_k,P_l\}=0 \;, \\
P_jP_k \propto I, \quad \{P_j,P_{l'}\}=\{P_k,P_{l'}\}=0\;,
\end{align}
for any non-identity $P_l, P_{l'}\in \mathcal{A}$.
\end{boxedresultdef} Result~\ref{result:MultiPauliPhase2} assumes nonzero fractional angles for \emph{all} commuting terms. The second line includes the identity-product sector, which corresponds to $j=k$ up to phase in the Pauli basis; unlike Results~\ref{result:OnePauli}--\ref{result:MultiPauli}, this compensated unequal-angle setting can therefore admit diagonal-sector first-order contributions. Under this most general assumption, it yields the maximal set of 36 dissipator elements capable of producing first-order fidelity asymmetry. Fig.~\ref{fig:dissipator_asymmetry_patterns}(c) shows the dissipator asymmetry pattern for a CZ gate constructed with dominant $ZZ$ interaction with $IZ$ and $ZI$ phase compensation. 

\subsection{Canonical physical dissipators: amplitude damping and pure dephasing}
\label{SubSec:PhysDiss}

Having identified which elements of $\boldsymbol{\beta}$ give rise to first-order fidelity asymmetry, an immediate question is how these contributions combine for physically relevant dissipators with a PSD $\boldsymbol{\beta}$ matrix. To address this, we analyze the standard single-qubit amplitude-damping and pure-dephasing dissipators (Appendix~\ref{App:AsymmSim}).

First, we observe that a single-qubit pure-dephasing dissipator of the form $\beta_{\phi,q}\mathcal{D}[Z_q]$, for either qubit $q = a,b$, produces only a second-order fidelity asymmetry. This is because none of the gate instances examined above (Fig.~\ref{fig:dissipator_asymmetry_patterns}) exhibit asymmetry associated with the $IZ$ or $ZI$ components of the $\boldsymbol{\beta}$ matrix. Second, a standard single-qubit $T_1$ dissipator of the form $\beta_{\downarrow,q}\mathcal{D}[S^-_q]$, with $S^-_q \equiv (X_q + iY_q)/2$, gives rise to four nonzero terms in the Pauli decomposition of Eq.~\eqref{eq:Def_of_Dbeta}. Specifically, the nonvanishing elements are $\beta_{X_qX_q} = \beta_{Y_qY_q} = \beta_{\downarrow}/4$ and $\beta_{X_qY_q} = \beta^*_{Y_qX_q} = -i\beta_{\downarrow}/4$. For the families of gates considered above, we find that the resulting first-order asymmetry contributions cancel identically (Appendix~\ref{App:AsymmSim}).    

This indicates that, beyond the selection rules identified above, additional first-order cancellations of the asymmetry can occur when the relevant $\beta$-matrix elements are balanced. The structure of the $\beta$ matrix depends on the microscopic noise model and on the derivation of the corresponding Lindblad-form master equation. Determining the general physical conditions under which the $\beta$ matrix does not induce such asymmetry cancellations therefore remains an important direction for future work.

%%%%%%%% Fig:Gauge-consistent learning %%%%%%%%%%%%%
\begin{figure*}
\centering
\includegraphics[scale=0.190]{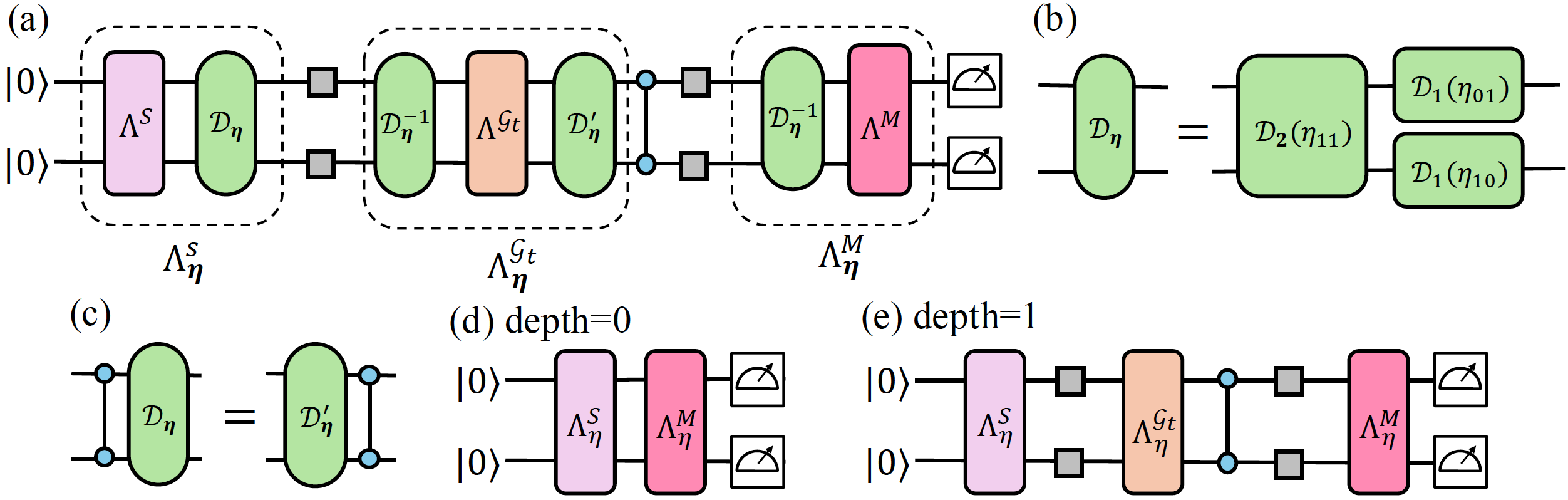}
\caption{\textbf{Physical self-consistent Pauli noise learning} -- (a) An example circuit for a two-qubit gate, with state preparation noise $\Lambda^{S}$, two-qubit (CZ) gate noise $\Lambda^{\mathcal{G}_{\text{t}}}$, and measurement noise $\Lambda^{M}$. Inserting Gauge channels $\mathcal{D}_{\eta}$ and its inverse dresses each noise channel without altering the logical action of the circuit. (b) Gauge degrees of freedom for Pauli channels can be modeled as a generalized depolarizing channel $\mathcal{D}_{\eta}$, decomposed further into single- and two-qubit depolarizing channels. (c) The conjugated depolarizing gauge channel emerges when pushed through the ideal (CZ) gate and is defined as $\mathcal{D}'_{\eta} \equiv \mathcal{U}_g^{-1} \mathcal{D}_{\eta} \mathcal{U}_g$. (d) Depth-0 learning circuit with the aggregate SPAM noise product $\Lambda_{\eta}^{M}\Lambda_{\eta}^{S}$. (e) Depth-1 learning circuit with the added gate noise. The physical (symmetric) self-consistent learning fixes $\eta$ by enforcing $\Lambda_{\eta}^{\mathcal{G}_{\text{t}}} = \mathcal{U}_g^{-1} \Lambda_{\eta}^{\mathcal{G}_{\text{t}}}\mathcal{U}_g$.} 
\label{fig:gauge_consistent_learning}
\end{figure*}
%%%%%%%%%%%%%%%%%%%%%%%%%%%%%%%%%%%%%%%%

%%%%%%%%%%%%%%% Sec: Physical Gauge-consistent learning %%%%%%%%%%%%%%%%%%
\section{Pauli noise learning in the symmetric gauge}
\label{Sec:GaugeLearning}

As an application of our approximate-symmetry results, we derive a gauge-fixing protocol that separates state-preparation and measurement errors in experiments. Consider a gate set consisting of initialization to
$\ket{0}^{\otimes n}$, computational-basis measurement,
single-qubit gates (with negligible or gate-independent
noise~\cite{wallman2016noise}), and entangling Clifford gate
$\mathcal{G}$. Under Pauli twirling, each noisy operation
acquires a Pauli noise channel:
$\tilde\rho_0 = \Lambda^S(\rho_0)$,
$\tilde{\mathcal{M}} = \mathcal{M}\Lambda^M$, and
$\mathcal{G}_{\text{t}} = \mathcal{U}_g\Lambda^{\mathcal{G}_{\text{t}}}$.
As shown in Refs.~\cite{chen2023learnability, chen2026efficient}
and Fig.~\ref{fig:gauge_consistent_learning}, a generalized depolarizing
channel $\mathcal{D}_{\bm\eta}$ can be inserted between these
channels without changing any experimental outcome, giving the
gauge transformation
\begin{align}
\Lambda^S &\mapsto \mathcal{D}_{\bm\eta}\Lambda^S, \quad
\Lambda^M \mapsto \Lambda^M\mathcal{D}_{\bm\eta}^{-1},
\nonumber\\
\Lambda^{\mathcal{G}_{\text{t}}} &\mapsto
\mathcal{D}'_{\bm\eta}\Lambda^{\mathcal{G}_{\text{t}}}
\mathcal{D}_{\bm\eta}^{-1},
\label{eq:gauge_transform}
\end{align}
where
$\mathcal{D}'_{\bm\eta} =
\mathcal{U}_g^{-1}\mathcal{D}_{\bm\eta}\mathcal{U}_g$.
Since $\mathcal{D}_{\bm\eta}$ commutes with single-qubit gates,
this transformation preserves all experimental outcomes and all
structural assumptions of the Pauli noise
model~\cite{chen2026efficient}.

In the self-consistent learning framework of Ref.~\cite{chen2026efficient}, one designs a set of experiments where each experiment consists of a sequence of Clifford gates and a Pauli observable measured at the end. The negative logarithm of each measured expectation value is a linear combination of the log-fidelity of the noise channels, yielding a linear system $\bm{y} = F\bm{x}$. Here, $\bm{y}$ is the vector of measured log-expectation values, $\bm{x}$ collects all log-fidelity of the noise channels in the gate set, and $F$ is a design matrix determined by the experiment design. The null space of $F$ corresponds to the gauge degrees of freedom. Under a quasi-local noise ansatz, the gauge reduces to $n$ single-qubit depolarizing parameters, each controlling how noise is partitioned between state preparation and measurement on a given qubit (Appendix~\ref{App:SymmetricGauge}).

Our symmetry analysis provides a way to fix these parameters. Results~\ref{result:OnePauli}--\ref{result:MultiPauliPhase2} establish that under common \textit{dominant} noise sources in quantum processors, $\Delta f_{mn} = O(\epsilon^2)$ where $\epsilon$ is the generic (coherent or dissipative) noise strength. Specifically, it implies that
\begin{align}
f_{P}^{\mathcal{G}_{\text{t}},\eta} =
f_{U_g P U_g^\dagger}^{\mathcal{G}_{\text{t}},\eta}
\quad \forall\; \text{conjugate pairs} \, (P, U_g P U_g^\dagger).
\label{eq:symmetric_gauge}
\end{align}
The symmetric condition~\eqref{eq:symmetric_gauge}, together with the assumption of local state-preparation and measurement errors, uniquely determines $\bm\eta$ (Appendix~\ref{App:ExptDet} and Fig.~\ref{fig:gauge_consistent_learning}). For a two-qubit CZ gate under the local gauge ansatz, the residual gauge freedom is captured by two single-qubit depolarizing parameters. Imposing the two CZ-pair symmetry conditions $f_{XI}=f_{XZ}$ and $f_{IX}=f_{ZX}$ then fixes the gauge parameters uniquely. 

The protocol proceeds in two stages. Depth-0 circuits (no entangling gates) measure the SPAM product
$\Lambda^M\Lambda^S$ for each Pauli basis (Fig.~\ref{fig:gauge_consistent_learning}(d)). Depth-1 circuits
add a single entangling gate layer, measuring the combined
SPAM-plus-gate fidelity
(Fig.~\ref{fig:gauge_consistent_learning}(e)). The symmetry
condition~\eqref{eq:symmetric_gauge} then provides the
constraints needed to decompose the depth-0 SPAM product into
individual state-preparation and measurement contributions,
since $\bm\eta$ determines how noise is split between
$\Lambda^S$ and $\Lambda^M$.

Our approach differs from existing methods in two respects. First, the TREX method~\cite{vandenberg2022model} assumes noiseless state preparation in order to calibrate measurement noise. In contrast, our method makes no assumptions about the magnitude of either SPAM component. Second, it relies only on the physically motivated condition that the dominant gate-noise mechanisms do not generate first-order fidelity asymmetry, a condition we derive from underlying physical principles rather than impose arbitrarily. Compared to the fully self-consistent approach of
Ref.~\cite{chen2025disambiguating}, which leaves the gauge free
and optionally optimizes it to reduce PEC sampling overhead, our
method actively fixes the gauge using device physics. The two
approaches are complementary. The symmetric gauge can be used only when the dominant noise mechanisms are understood.

We further benchmark our symmetric-gauge protocol against the recent ancilla-assisted SPAM characterization protocols of Ref.~\cite{yu2025efficient}. Those protocols assume a noise structure that depends on the direction of the applied gates, an assumption that does not necessarily hold in practice. As detailed in Appendix~\ref{app:comparison} and summarized in Fig.~\ref{fig:comparison}, under realistic gate implementations and physically motivated noise models, our protocol remains accurate across the full range of noise parameter values, whereas the protocols of Ref.~\cite{yu2025efficient} show significant deviations in the inferred SPAM parameters. 

%% ============================================================
%% Section: Experiments
%% ============================================================
\section{Experiments}
\label{Sec:Experiment}
We validate the symmetric-gauge learning protocol on IBM Kingston using CZ entangling gates with IZ/ZI phase compensation. Sixteen non-overlapping qubit pairs are driven simultaneously in a
single parallel circuit of width 32~qubits; full details of the
circuit construction and data processing are given in
Appendix~\ref{App:ExptDet}.

The protocol follows Fig.~\ref{fig:gauge_consistent_learning}.
A depth-0 circuit prepares and measures all qubits in the
computational ($Z$) basis, yielding the SPAM fidelity products
$f^S_P f^M_P$ for $P \in \{IZ, ZI, ZZ\}$.  Two depth-1
circuits conjugate a CZ layer by Hadamard gates on, respectively,
the first or the second qubit of every pair, giving access to the
gate Pauli fidelities $f^{\mathcal{G}_{\text{t}}}_{XI}$ and
$f^{\mathcal{G}_{\text{t}}}_{IX}$ as well as their CZ conjugates
$f^{\mathcal{G}_{\text{t}}}_{XZ}$ and $f^{\mathcal{G}_{\text{t}}}_{ZX}$.
The resulting system of equations is under-determined and must be
closed by a gauge choice.  In the \emph{default (CZ-symmetric)
gauge}, the conditions
$f^{\mathcal{G}_{\text{t}}}_{XI} = f^{\mathcal{G}_{\text{t}}}_{XZ}$ and
$f^{\mathcal{G}_{\text{t}}}_{IX} = f^{\mathcal{G}_{\text{t}}}_{ZX}$, together with single-qubit locality of the SPAM channels, uniquely determine the SPAM decomposition.  We also consider an \emph{SP-perfect gauge} that instead fixes $f^S_{IZ} = f^S_{ZI} = 1$, attributing all error to measurement
and gate channels.  All circuits are Pauli-twirled with 500 randomizations at 100~shots each.

To illustrate how different choices of gauge assign errors to individual components, we inject synthetic incoherent state-preparation
errors.  For a given qubit pair, we collect data under two state-preparation settings. In the first case, we simply run the protocol as described, and in the second case a bit-flip is applied to
the first qubit. By linearity, we emulate an additional bit-flip error channel with probability $p$ on the first qubit by forming a convex combination of the measured expectation values for each Pauli observable $\langle P\rangle$:
\begin{equation}
  \langle P\rangle_{\mathrm{synth}}(p)
  = (1-p)\,\langle P\rangle_{I}
  + p\,\langle P\rangle_{X},
  \label{eq:synth_mixing}
\end{equation}
where the subscripts $I$ and $X$ denote the data obtained without and with the injected bit flip, respectively. Because the noise acts solely on state
preparation, a faithful SPAM-gate separation must attribute the
resulting fidelity loss entirely to the state-preparation channel
while leaving measurement and gate fidelities unchanged.

\begin{figure}[t]
  \centering
  \includegraphics[width=\columnwidth]{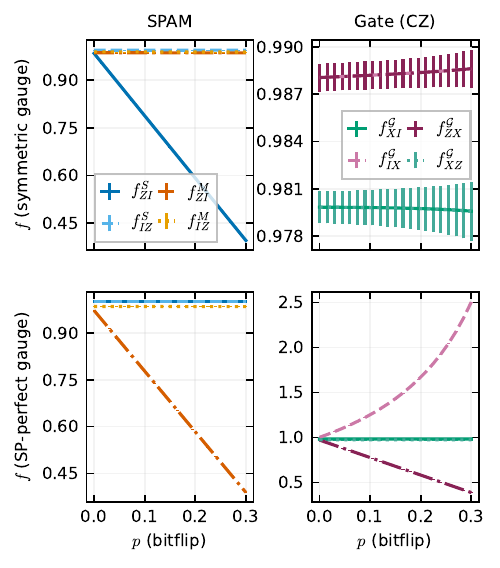}
  \caption{Extracted fidelities versus synthetic bit-flip
    strength~$p$ (X applied to $q_4$) for qubit pair $(4,5)$.
    The CZ-symmetric gauge is shown at the top and
    SP-perfect gauge is shown at the bottom.
    Left panels show state-preparation and measurement fidelities;
    right panels show gate (CZ) fidelities.  In the symmetric gauge,
    only SP fidelities decrease with~$p$; in the SP-perfect gauge,
    the error is misattributed to measurement and gate channels.}
  \label{fig:synth_validation}
\end{figure}

Figure~\ref{fig:synth_validation} shows the extracted fidelities versus mixing strength~$p$ for a representative qubit pair in
both gauges. Under the symmetric gauge (top row), increasing $p$ reduces only the state-preparation fidelity $f^S_{ZI}$, while measurement and CZ gate fidelities remain flat
within statistical uncertainty. In the SP-perfect gauge (bottom row), state preparation is forced to unit fidelity by construction; the
injected error leaks into the measurement and gate channels, with some gate fidelities diverging well above unity.  This contrast
demonstrates that the choice of gauge is physically meaningful and that the symmetric gauge provides a more consistent attribution of the injected error. We emphasize, however, that this observation alone does not establish the symmetric gauge as the true gauge. Any intrinsic asymmetry still contributes to state-preparation and measurement errors, but this
contribution appears as a constant offset at~$p=0$. Thus, in the symmetric gauge, the error in our estimate of the state-preparation error is bounded by the magnitude of this asymmetry.

\begin{figure}[t]
  \centering
  \includegraphics[scale=0.92]{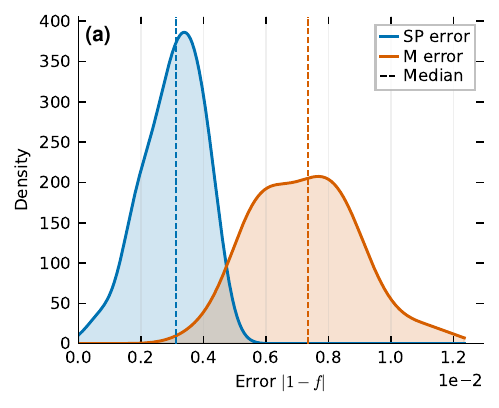}
    \includegraphics[scale=0.92]{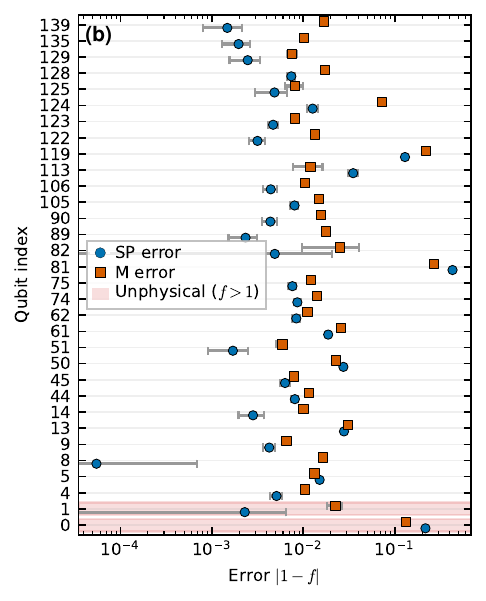}
  \caption{Per-qubit SPAM errors at $p=0$ across all 32~qubits
    of the 16 parallel CZ pairs on IBM Kingston. (a) Kernel-density estimates of the state-preparation (SP) and measurement (M) error distributions, with dashed vertical lines marking the corresponding medians.
(b) Per-qubit SPAM errors. Circles denote state-preparation errors and squares denote measurement errors. Red bands mark qubit pairs for which any extracted fidelity exceeds unity, indicating an unphysical estimate.}
  \label{fig:spam_errors}
\end{figure}

We next run the protocol, without noise injection, to estimate SP and M errors separately. As a self-consistency check we verify the physicality of the extracted fidelities: all Pauli fidelities must satisfy $f \le 1$.  Violations indicate that statistical noise or model assumptions have pushed the solution outside the physical domain, flagging the corresponding qubit pair as unreliable. Figure~\ref{fig:spam_errors} displays the per-qubit SPAM errors ($|1 - f^S_P|$ and $|1 - f^M_P|$) at $p = 0$ across all 32~qubits on a logarithmic scale. As shown in panel (a), measurement errors are systematically larger than state-preparation errors across the device, consistent with the known readout asymmetry of superconducting transmon qubits. Panel (b) shows the corresponding SPAM breakdown for each qubit. Red bands highlight qubit pairs in which any fidelity, i.e., state preparation, measurement, or gate, exceeds unity. Pairs without red bands pass the physicality check, giving confidence in our method for separating SPAM errors. 

\section{Conclusion}
\label{Sec:Conclusion}
We showed that physically motivated Lindbladian structure imposes approximate symmetry constraints on Pauli fidelities of two‑qubit Clifford gates, enabling principled gauge fixing for SPAM without assuming error magnitudes. At first order, coherent errors do not induce fidelity asymmetry, whereas the symmetry can be broken only by a restricted class of dissipator elements, most of which are off-diagonal in the Pauli basis. Perturbative analysis, simulation, and experiments on IBM Kingston support these predictions, justifying imposing first‑order fidelity symmetry to set the gauge and separate SPAM with depth‑0/1 experiments.

The protocol as developed applies to individual gates characterized in isolation, so with this choice of gauge, SPAM errors can be separated from gate errors one gate at a time. Using this result in large-scale applications therefore requires two steps. First the protocol is applied gate by gate to fix the gauge and extract the separated state preparation and measurement errors, and then these inferred errors can be used to perform SPAM-robust gate characterization within scalable Pauli-channel learning (e.g., cycle benchmarking \cite{erhard2019characterizing, flammia2020efficient}, SPL models \cite{vandenberg2023probabilistic, berg2023techniques}) and model-based error-mitigation methods~\cite{temme2017error, vandenberg2023probabilistic, kim2023evidence, filippov2023scalable}. Our results also reveal the dissipator elements that can cause leading-order asymmetry and inform hardware-aware calibrations to suppress them.

A more direct route would be to derive the analogous symmetry constraints for a full layer of parallel two-qubit gates rather than for gates addressed individually. Because scalable characterization and mitigation protocols operate on such layers directly, a layer-level symmetry would let SPAM be gauge-fixed and separated within the large-scale protocol itself, rather than as a separate gate-by-gate step performed beforehand. This raises the question of how crosstalk and simultaneous dissipative processes across a layer combine to preserve or break the fidelity symmetry we identified at the single-gate level, and whether the same restricted class of off-diagonal dissipator elements remains the relevant first-order terms when gates act in parallel. We leave this extension to future work.

\section{Acknowledgments}
\label{Sec:Acknowledge}
We acknowledge helpful discussions with Ewout van den Berg, Abhinav Kandala, and Senrui Chen.  
%%%%%%%%%%%%%%%%%%%%%%%%%%%%%%%%%%%%%%%%%%%%%%%%%%%%%%%%%%%%%
\clearpage

\appendix
%%%%%%%%%%% Appendix: Lindblad perturbation %%%%%%%%%%%%%%%%%%%%%%%%%%
\section{Pauli noise asymmetry via Lindblad perturbation}
\label{App:LindPert}
In this Appendix, we describe our procedure for deriving Pauli‑noise symmetries~\cite{chen2023learnability} from continuous‑time Lindblad noise models~\cite{Blume_Taxonomy_2022, malekakhlagh2025efficient, miller2025efficient, berg2025large}. Our analysis is based on time‑dependent Lindblad perturbation theory~\cite{Dai_Floquet_2016, Schnell_High_2021, Malekakhlagh_Time_2022, Mizuta_Breakdown_2021, Ture_Application_2024, Huang_Towards_2025, malekakhlagh2025efficient}, which serves as the open‑system analogue of the widely used Magnus expansion~\cite{Magnus_Exponential_1954, Blanes_Magnus_2009, Blanes_Pedagogical_2010, Puzzuoli_Algorithms_2023} and Dyson series~\cite{Dyson_SMatrixQED_1949, Dyson_Radiation_1949, Shillito_Fast_2021, Puzzuoli_Algorithms_2023}. Our notation and conventions align closely with Ref.~\cite{malekakhlagh2025efficient}. We subsequently apply the method to Clifford gates generated by a single Pauli and by sums of commuting Paulis in Appendices~\ref{App:OnePauliGates}--~\ref{App:MultiPauliWithPhase2}. 

We model a noisy two-qubit gate in terms of the following Lindblad equation
\begin{align}
\dot{\rho}(t) = -i [H_g + H_{\delta}(t), \rho(t)] + \mathcal{D}_{\beta}(t)\rho(t) \;,
\end{align}
 where $H_g$, $H_{\delta}$ and $\mathcal{D}_{\beta}$ denote the ideal Hamiltonian, noise Hamiltonian, and incoherent noise, respectively. More explicitly, in the Pauli basis, we have
\begin{align}
 \dot{\rho}(t) = - i [H_g +\sum\limits_j \frac{\delta_j}{2} P_j,\rho(t)] + \sum\limits_{jk} \beta_{jk} \mathcal{D}[P_j,P_k]\rho(t)  \;,
\end{align}
assuming arbitrary Hamiltonian noise with strength $\delta_j$, and arbitrary dissipator noise defined as
\begin{align}
& \mathcal{D}[P_j,P_k]\rho(t) \equiv P_j \rho(t) P_k^{\dag} - \frac{1}{2} \{P_k^{\dag} P_j,\rho(t) \}\;,
\end{align}
with a positive semi-definite dissipative matrix $\bm{\beta}$.

To extract the gate noise channel, it is helpful to first transform to the interaction frame with respect to $H_g$ as
\begin{align}
P_j(t) & \equiv U_g^{\dag}(t) P_j U_g(t) \;,\\
\rho_{I}(t) & \equiv U_g^{\dag} \rho(t) U_g(t) \;,
\end{align}
where $U_g(t)\equiv \exp(-i H_g t)$, resulting in dynamics in terms of noise only:
\begin{align}
\begin{split}
\dot{\rho}_I(t) = \mathcal{L}_I(t) \rho_I(t) &= - i [\sum\limits_j \delta_j P_j(t),\rho_I(t)] \\
&+ \sum\limits_{jk} \beta_{jk} \mathcal{D}[P_j(t),P_k(t)]\rho_I(t)  \;.
\end{split}
\end{align}
Then, the gate noise channel can be approximated under Lindblad-Dyson perturbation theory as
\begin{align}
\Lambda^{\mathcal{G}} \equiv e^{\Omega(\tau_g,0)} \approx \mathcal{I} + \Omega_1(\tau_g,0) + \Omega_2(\tau_g,0) + \frac{1}{2}\Omega_1^2(\tau_g,0) \;,  
\end{align}
with the leading-order generators depending on $\mathcal{L}_I$ according to 
\begin{align}
& \Omega_1(\tau_g,0) \equiv \int_{0}^{\tau_g}dt' \mathcal{L}_I(t') \;,  \\
& \Omega_2(\tau_g,0) \equiv \frac{1}{2}\int_{0}^{\tau_g} dt' \int_{0}^{t'} dt'' [\mathcal{L}_I(t'),\mathcal{L}_I(t'')] \;.
\end{align}
The noise of Clifford gates is commonly twirled in practice into an effective Pauli noise channel using randomized compiling \cite{wallman2016noise,hashim2021randomized} as
\begin{align}
\Lambda^{\mathcal{G}_t}\equiv \mathcal{T}_{P}(\Lambda^{\mathcal{G}}) = \frac{1}{|\mathcal{P}_N|}\sum\limits_{\mathcal{P}\in \mathcal{P}_N} \mathcal{P}^{\dag}\Lambda^{\mathcal{G}}\mathcal{P} \;.
\end{align}

A useful diagnostic is the fidelity asymmetry between Pauli operators related by the gate action, defined as
\begin{align}
\Delta f_{mn} \equiv 
\frac{1}{D}\Tr\!\left[P_m^{\dagger}\Lambda^{\mathcal{G}_{\text{t}}}(P_m)\right]
-\frac{1}{D}\Tr\!\left[P_n^{\dagger}\Lambda^{\mathcal{G}_{\text{t}}}(P_n)\right],
\label{Eq:LindPert-DefOfDelfmn}
\end{align}
where $P_n = U_g^{\dagger} P_m U_g$, $U_g \equiv U_g(\tau_g) = e^{-i H_g \tau_g}$, and $D=2^N$ is the Hilbert-space dimension.
%%%%%%%%%%%%%%%%%%%%%%%%%%%%%%%%%%%%%%%%%%%%%%%%%%%%%%%%%%%%%%%%%%%%%%%%

%%%%%%%%%%%%%%%%%%%%%%%% Appendix: Gates generated by a single Pauli %%%%%%
\widetext
\section{Gates generated by a single Pauli term}
\label{App:OnePauliGates}
Here, we analyze gates generated by a single-Pauli Hamiltonian $H_g = (\omega_g/2)P_g$, where $\omega_g$ is the physical gate coupling strength. We evaluate the fidelity asymmetry in Eq.~\eqref{Eq:LindPert-DefOfDelfmn} under Hamiltonian noise, diagonal dissipator noise, and off-diagonal dissipator noise, leading directly to Result~1 in the main text.    

The interaction-frame Pauli operator $P_{m}(t)$ has the explicit form
\begin{align}
P_{m}(t)\equiv e^{iH_g t} P_m e^{-iH_g t} = 
\begin{cases}
P_m, & [P_m, P_g] =0 \\
\cos\theta(t) P_m + i \sin\theta(t) P_g P_m, &\{P_m, P_g\} = 0
\end{cases}
\label{Eq:OnePauliGates-Def of P_I(t)}
\end{align}
where $\theta(t)=\omega_g t$. At Clifford angle $\theta(t)=\pi/2$ one finds
\begin{align}
P_n = U_g^{\dag} P_m U_g = i P_g P_m \;.
\end{align}
Using this, the asymmetry measure can be simplified to
\begin{align}
\Delta f_{mn} \equiv \frac{1}{D}\Tr\{P_m^{\dag}\Lambda^{\mathcal{G}_{\text{t}}}(P_m)\} -\frac{1}{D}\Tr\{P_m^{\dag}P_g^{\dag}\Lambda^{\mathcal{G}_{\text{t}}}(P_gP_m)\} \;.
\label{Eq:LindPert-DefOfDelfmn2}
\end{align}

\subsection{Hamiltonian noise}
Assume the noise consists of a single Hamiltonian term as $H_{\delta}= (\delta_l/2) P_l$.  The first-order fidelity asymmetry reads
\begin{align}
\begin{split}
\Delta f_{mn}^{(1)} &= \frac{1}{D}\int_0^{\tau_g} dt' \Tr\{-i\frac{\delta_l}{2} P_m^{\dag}[P_{l}(t'),P_m]\} \\
&- \frac{1}{D}\int_0^{\tau_g} dt' \Tr\{-i\frac{\delta_l}{2} P_m^{\dag}P_g^{\dag}[P_{l}(t'),P_gP_m]\} \;.
\end{split}
\end{align}
Here, each individual term is zero as the trace of a commutator is zero due to the cyclic property. The first commutator can be simplified as
\begin{align}
\Tr\{P_m^{\dag}[P_{l}(t'),P_m]\} = \Tr\{P_m^{\dag}P_{l}(t')P_m \} - \Tr\{P_m^{\dag}P_mP_{l}(t') \}= 0 \;.
\end{align}
The second commutator simplifies to:
\begin{align}
\begin{split}
\Tr\{P_m^{\dag}P_g^{\dag}[P_{l}(t'),P_gP_m]\} & = \Tr\{P_m^{\dag}P_g^{\dag}P_{l}(t')P_gP_m\} \\ 
& - \Tr\{P_m^{\dag}P_g^{\dag}P_gP_mP_{l}(t') \}= 0 \;.
\end{split}
\end{align}

It can be shown that second‑order contributions from Hamiltonian noise are generally nonzero; however, we do not analyze them further as they only constitute small corrections.

\subsection{Diagonal dissipator}
Consider a single diagonal dissipator noise term as $\mathcal{D}_\beta = \beta_{ll} (P_l \bullet P_l^{\dag} - I\bullet I)$. The first-order fidelity asymmetry is found as
\begin{align}
\begin{split}
\Delta f_{mn}^{(1)} &= \frac{1}{D}\int_0^{\tau_g} dt' \Tr\{\beta_{ll} P_m^{\dag}[P_{l}(t')P_mP_{l}^{\dag}(t')-P_m]\} \\
&- \frac{1}{D}\int_0^{\tau_g} dt' \Tr\{\beta_{ll} P_m^{\dag}P_g^{\dag}[P_{l}(t')P_gP_mP_{l}^{\dag}(t')-P_gP_m]\} \;.
\end{split}
\label{Eq:OnePauliGates-Delf_mn diag_D 1}
\end{align}
Note that the second terms in the two lines cancel each other since $\Tr\{P_m^{\dag}P_m\} = \Tr\{P_m^{\dag}P_g^{\dag}P_gP_m\} =\Tr\{I\}$. To understand the interplay of first terms, we consider two cases:
\begin{itemize}
\item Case (i) -- $[P_l,P_g] =0$ \\
Here, based on Eq.~\eqref{Eq:OnePauliGates-Def of P_I(t)}, we find that $P_l(t)=P_l$. Therefore, the asymmetry is zero since both terms simplify to $\Tr\{P_m^{\dag}P_g^{\dag}P_{l}(t')P_gP_mP_{l}^{\dag}(t')\} = \Tr\{P_m^{\dag}P_lP_mP_l^{\dag}\}$.

\item Case (ii) -- $\{P_l,P_g\} = 0$ \\
Here, we first rewrite the asymmetry as
\begin{align}
\begin{split}
\Delta f_{mn}^{(1)} & = \frac{1}{D}\int_{0}^{\tau_g} dt' \beta_{ll} \Tr\{P_m^{\dag}[P_{l}(t')-P_g^{\dag}P_{l}(t')P_g]P_m P_{l}^{\dag}(t')\} \\
& = \frac{2}{D}\int_{0}^{\tau_g} dt' \beta_{ll} \Tr\{P_m^{\dag}\Pi_{P_g}^{-}[P_{l}(t')]P_m P_{l}^{\dag}(t')\}\;,
\end{split}
\label{Eq:OnePauliGates-Delf_mn diag_D 2}
\end{align}
$\Pi_{P_g}^-(X)\equiv\tfrac{1}{2}(X-P_g^\dagger X P_g)$ is the projector onto the \emph{odd} part under $P_g$. Given that $\{P_l,P_g\} = 0$, we have $P_{l}(t) = \cos\theta(t) P_l + i\sin\theta(t) P_g P_l$, using which we find $P_g^{\dag}P_{l}(t)P_g =-P_{l}(t)$. The asymmetry $\Delta f_{mn}^{(1)}$ simplifies to:
\begin{align}
\begin{split}
\Delta f_{mn}^{(1)} & =  \frac{2}{D}\int_{0}^{\tau_g} dt' \beta_{ll} \Tr\{P_m^{\dag}P_{l}(t')P_m P_{l}^{\dag}(t')\} \\
& = 2 \frac{\beta_{ll}}{D} \Tr\{P_m^{\dag}P_lP_mP_l^{\dag}\}\int_{0}^{\tau_g} dt' \cos^2 \theta(t') \\
& +2 \frac{\beta_{ll}}{D} \Tr\{P_m^{\dag}P_gP_lP_mP_l^{\dag}P_g^{\dag}\}\int_{0}^{\tau_g} dt' \sin^2 \theta(t') \\
& -2i \frac{\beta_{ll}}{D} \Tr\{P_m^{\dag}P_lP_mP_l^{\dag}P_g^{\dag}\}\int_{0}^{\tau_g} dt' \sin \theta(t') \cos \theta (t') \\
& +2i \frac{\beta_{ll}}{D} \Tr\{P_m^{\dag}P_gP_lP_mP_l^{\dag}\}\int_{0}^{\tau_g} dt' \sin \theta(t') \cos \theta (t') \;.
\label{Eq:OnePauliGates-Delf_mn diag_D 3}
\end{split}
\end{align}

All integrals are in principle non-zero. In particular, assuming constant-amplitude square pulses at the Clifford angle $\omega_g\tau_g=\pi/2$, we have 
\begin{align}
&\int_{0}^{\tau_g} dt' \cos^2 \theta (t') = \int_{0}^{\tau_g} dt' \sin^2 \theta (t') = \frac{\pi}{4\omega_g} \;, \\
&\int_{0}^{\tau_g} dt' \sin \theta(t') \cos \theta (t') = \frac{1}{2\omega_g} \;. 
\end{align}

We show Eq.~\eqref{Eq:OnePauliGates-Delf_mn diag_D 3} is zero for the square-pulse case using commutativity and trace properties of Pauli operators. First, by the fact we are looking for asymmetry of $P_m$, and its pair under a gate generated by $P_g$, they must anti-commute $\{P_g,P_m\}=0$. Using this, we can rewrite the trace on the 2nd line of Eq.~\eqref{Eq:OnePauliGates-Delf_mn diag_D 3} as 
\begin{align}
\begin{split}
\Tr\{P_m^{\dag}P_gP_lP_mP_l^{\dag}P_g^{\dag}\} & = \Tr\{P_m^{\dag}P_lP_gP_mP_g^{\dag}P_l^{\dag}\} \\
& = - \Tr\{P_m^{\dag}P_lP_mP_l^{\dag}\} \;,
\end{split}
\end{align}
where first we pushed $P_l$ through $P_g$ and lastly we used $P_gP_mP_g^{\dag}=-P_m$. Therefore, the first two lines of Eq.~\eqref{Eq:OnePauliGates-Delf_mn diag_D 3} cancel each other (under the condition $\int_{0}^{\tau_g} dt' \cos^2 \theta (t') = \int_{0}^{\tau_g} dt' \sin^2 \theta (t')$) . The last two terms of Eq.~\eqref{Eq:OnePauliGates-Delf_mn diag_D 3} are individually zero given that they simplify to the trace of a non-identity Pauli operator. 
\end{itemize}

\subsection{Off-diagonal dissipator noise}
Next, we assume an off-diagonal dissipator noise term of the form 
\begin{align}
\begin{split}
\mathcal{D}_\beta & = \beta_{jk} \mathcal{D}[P_j,P_k] = \beta_{jk} (P_j \bullet P_k^{\dag} - \frac{1}{2}\{P_k^{\dag}P_j, \bullet \}) \;, 
\end{split}
\label{Eq:OnePauliGates-OffDiagD}
\end{align}
where $j\neq k$. The fidelity asymmetry for Pauli pairs $P_m$ and $P_n$ that are mapped by the ideal gate according to $P_n = iP_g P_m$ can be written as
\begin{align}
\begin{split}
\Delta f_{mn,jk}^{(1)} &= \frac{1}{D} \int_0^{\tau_g} dt' \Tr\{\beta_{jk} P_m^{\dag}[P_j(t')P_mP_{k}^{\dag}(t')-\frac{1}{2}\{P_{k}^{\dag}(t')P_j(t'),P_m\}]\} \\
&- \frac{1}{D}\int_0^{\tau_g} dt' \Tr\{\beta_{jk} P_m^{\dag}P_g^{\dag}[P_j(t')P_gP_mP_{k}^{\dag}(t')-\frac{1}{2}\{P_{k}^{\dag}(t')P_j(t'),P_gP_m \}]\} \;.
\end{split}
\label{Eq:OnePauliGates-OffDiagD1}
\end{align}

We again find that the anticommutator (second) terms of each line cancel each other since
\begin{align}
& \Tr \{-\frac{1}{2}P_m^{\dag} \{P_{k}^{\dag}(t')P_j(t'),P_m\}\}  = -\Tr \{P_{k}^{\dag}(t')P_j(t')\} \\
& \Tr \{-\frac{1}{2}P_m^{\dag}P_g^{\dag} \{P_{k}^{\dag}(t')P_j(t'),P_gP_m\}\}  = -\Tr \{P_{k}^{\dag}(t')P_j(t')\}
\end{align}
Therefore, we can simplify $\Delta f_{mn,jk}^{(1)}$ as
\begin{align}
\Delta f_{mn,jk}^{(1)} = \frac{2}{D}\int_{0}^{\tau_g} dt' \beta_{jk} \Tr\{P_m^{\dag}\Pi_{P_g}^{-}[P_j(t')]P_m P_{k}^{\dag}(t')\} \;.
\label{Eq:OnePauliGates-OffDiagD2}
\end{align}

Depending on the commutativity of $P_g$ and $P_j$ there are two cases:
\begin{itemize}
\item Case (i) -- $[P_g,P_j]=0$ \\
Here, we also find that $[P_g,P_j(t)]=0$ and therefore $\Delta f_{mn,jk}^{(1)}=0$

\item Case (ii) -- $\{P_g,P_j\}=0$\\
Here, we find $P_g^{\dag}P_j(t')P_g=-P_j(t')$ and we can write $\Delta f_{mn,jk}^{(1)}$ as
\begin{align}
\begin{split}
\Delta f_{mn,jk}^{(1)} & =  \frac{2}{D}\int_{0}^{\tau_g} dt' \beta_{jk} \Tr\{P_m^{\dag}P_j(t')P_m P_{k}^{\dag}(t')\} \\
& = 2 \frac{\beta_{jk}}{D} \Tr\{P_m^{\dag}P_jP_mP_k^{\dag}\}\int_{0}^{\tau_g} dt' \cos^2 \theta(t') \\
& +2 \frac{\beta_{jk}}{D} \Tr\{P_m^{\dag}P_gP_jP_mP_k^{\dag}P_g^{\dag}\}\int_{0}^{\tau_g} dt' \sin^2 \theta(t') \\
& -2i \frac{\beta_{jk}}{D} \Tr\{P_m^{\dag}P_jP_mP_k^{\dag}P_g^{\dag}\}\int_{0}^{\tau_g} dt' \sin \theta(t') \cos \theta (t') \\
& +2i \frac{\beta_{jk}}{D} \Tr\{P_m^{\dag}P_gP_jP_mP_k^{\dag}\}\int_{0}^{\tau_g} dt' \sin \theta(t') \cos \theta (t') \;.
\label{Eq:OnePauliGates-OffDiagD3}
\end{split}
\end{align}

In contrast to the diagonal case, here the first two terms of Eq.~\eqref{Eq:OnePauliGates-OffDiagD3} are zero as they simplify to the trace of a non-identity Pauli operator. For instance, the trace on the second line can be rewritten as:
\begin{align}
\begin{split}
\Tr\{P_m^{\dag}P_gP_jP_mP_k^{\dag}P_g^{\dag}\} &= \Tr\{P_g^{\dag}P_m^{\dag}P_gP_jP_mP_k^{\dag}\}  \\
& = -\Tr\{P_m^{\dag}P_jP_mP_k^{\dag}\} = \pm \Tr \{P_j P_k^{\dag}\}\;.    
\end{split}
\end{align}
Given that $j\neq k$, the trace is zero. The same applies to the first line of Eq.~\eqref{Eq:OnePauliGates-OffDiagD3}.

The last two terms of Eq.~\eqref{Eq:OnePauliGates-OffDiagD3} can be further simplified. In particular, the traces can be shown to be negatives of each other: 
\begin{align}
& \text{3rd line:} \ \Tr\{P_m^{\dag}P_jP_mP_k^{\dag}P_g^{\dag}\} = \Tr\{P_g^{\dag}P_m^{\dag}P_jP_mP_k^{\dag}\} \;,\\ 
& \text{4th line:} \  \Tr\{P_m^{\dag}P_gP_jP_mP_k^{\dag}\} = - \Tr\{P_g^{\dag}P_m^{\dag}P_jP_mP_k^{\dag}\} \;.
\end{align}
Therefore, the asymmetry can be expressed compactly as
\begin{align}
\Delta f_{mn,jk}^{(1)} = - 2 i \frac{\beta_{jk}}{D} \Tr\{P_g^{\dag}P_m^{\dag}P_jP_mP_k^{\dag}\}\int_{0}^{\tau_g} dt' \sin[2 \theta(t')]
\label{Eq:Delf_mn offdiag_D Sol}
\end{align}
\end{itemize}

The question that remains is under what condition the fidelity asymmetry is nonzero for a given off-diagonal dissipator. The trace in Eq.~\eqref{Eq:Delf_mn offdiag_D Sol} can be further simplified to $\Tr\{P_g^{\dag}P_m^{\dag}P_jP_mP_k^{\dag}\} = \pm \Tr\{P_gP_jP_k\}$. This suggests that the only combination that leads to a non-zero trace is when $P_jP_k \propto P_g$. In summary, for those off-diagonal elements whose product equals the gate generator $P_jP_k \propto P_g$, and anticommute as $\{P_j,P_g\}=\{P_k,P_g\}=0$, we expect a first-order contribution to the asymmetry, leading to our Result~\ref{result:OnePauli} in the main text.  

%%%%%%%%%%%%%%%%%%%%%%%%%%%%%%%%%%%%%%%%%%%%%%%%%%%%%%%%%%%%%%%%%%%

%%%%%%%%%%%%%%%%%% Appendix: Gates generated by common-angle commuting Pauli terms %%%%%%%%%
\section{Gates generated by common-angle commuting Pauli terms}
\label{App:MultiPauliGates}
In this Appendix, we consider two-qubit Clifford gates generated by sums of mutually commuting Pauli operators. Representative examples include CNOT, CZ, iSWAP, and SWAP, which (up to a global phase) can be written as 
\begin{align}
&\text{CNOT}=\exp[-i(\pi/4)(II+ZI+IX-ZX)]\;,\\ 
&\text{CZ}=\exp[-i(\pi/4)(II-IZ-ZI+ZZ)]\;,\\ 
&\text{iSWAP}= \exp[-i(\pi/4)(XX+YY)]\;, \\
&\text{SWAP}= \exp[-i(\pi/4)(XX+YY+ZZ)]\;.
\end{align}
In general, such gates involve at most three independent non-identity Pauli generators, denoted $G=\{P_a, P_b, P_c\}$, where $P_c = s_{ab}\, P_aP_b$ with sign $s_{ab}\in \{0, \pm1\}$. We show that Result~\ref{result:OnePauli} applies independently to each generator in $G$, yielding Result~\ref{result:MultiPauli} of the main text.   

We use a generic Hamiltonian for this family of gates as
\begin{align}
H_g = \frac{\omega_{g}}{2}(P_a+P_b+P_c) \;,
\label{Eq:MultiPauliGates-DefOfHg}
\end{align}
resulting in the unitary gate operator
\begin{align}
U_g(t) = e^{-i\frac{\theta(t)}{2}P_a}\;e^{-i\frac{\theta(t)}{2}P_b}\;e^{-i\frac{\theta(t)}{2}P_c},
\label{Eq:MultiPauliGates-DefOfUg}
\end{align}
where $\theta(t)=\omega_g t$, and we dropped any $II$ term as it only contributes a global phase.

We next derive an explicit expression for the interaction-frame Pauli operators, defined as
\begin{align}
P_j(t)\equiv U_g^\dagger(t)\,P_j\,U_g(t),
\end{align}
by sequentially applying the single-generator rotation identity
$e^{+i\frac{\theta}{2}P_G} P_j e^{-i\frac{\theta}{2}P_G}
=\cos\theta\,P_j+i\sin\theta\,P_G P_j$ when $\{P_G,P_j\}=0$ and leaving $P_j$ invariant when $[P_G,P_j]=0$ for $P_G \in G$. Let $c(t)\equiv \cos(\theta(t))$ and $s(t)\equiv \sin(\theta(t))$. Moreover, we define commutation signature bits
\begin{align}
b_{jk}\equiv\begin{cases}
0,& [P_j,P_k]=0 \;,\\
1,& \{P_j,P_k\}=0\;.
\end{cases}
\end{align}
Consequently, one obtains the closed-form identity (assuming $s_{ab}\neq 0$)
\begin{align}
P_j(t)=
\begin{cases}
P_j, & (b_{aj},b_{bj})=(0,0),\\[3pt]
c^2 P_j + i\,c s\, P_a P_j - s_{ab}\, s^2 P_b P_j + i\,c s\, P_c P_j, & (b_{aj},b_{bj})=(1,0),\\[3pt]
c^2 P_j - s_{ab}\, s^2 P_a P_j + i\,c s\, P_b P_j + i\,c s\, P_c P_j, & (b_{aj},b_{bj})=(0,1),\\[3pt]
c^2 P_j + i\,c s\,(P_a P_j+P_b P_j) - s_{ab}\, s^2\, P_c P_j, & (b_{aj},b_{bj})=(1,1),
\end{cases}
\label{Eq:MultiPauliGates-4termDecomp}
\end{align}
where $P_c=s_{ab}P_aP_b$ commutes with $P_j$ iff $b_{aj}=b_{bj}$, and anticommutes iff $b_{aj}\neq b_{bj}$. An important observation is that at Clifford angle $\theta(t)=\pi/2$, Pauli operators transform to one another only via a single generator term, leading to independent subspaces.

Following Eq.~\eqref{Eq:LindPert-DefOfDelfmn}, we define the asymmetry for a Pauli pair $(P_m,P_n)$ related by the ideal gate. Importantly, for the family of gates considered in Eqs~(\ref{Eq:MultiPauliGates-DefOfHg})--(\ref{Eq:MultiPauliGates-DefOfUg}), at Clifford angle $\theta=\pi/2$, the Pauli conjugation simplifies to multiplication by only one of the commuting generators
$P_l\in\{P_a,P_b,P_c\}$ (set $s=1$, $c=0$ in Eq.~\eqref{Eq:MultiPauliGates-4termDecomp}):
\begin{align}
P_n = U_g^\dagger\,P_m\,U_g = -s_{ab}\,P_l\,P_m \;,
\label{Eq:MultiPauliGates-CliffordPairing}
\end{align}
where $P_l$ appears as the product of the two generator terms that anticommute with $P_m$ and $P_n$, hence $[P_m,P_l]=[P_n,P_l]=0$. If $P_m$ commutes with all of $P_a,P_b,P_c$, then $P_n=P_m$ and the asymmetry is trivially zero.

For such family of gates, neither Hamiltonian noise nor diagonal dissipators produce a first‑order asymmetry. In the following, we derive the specific off‑diagonal dissipator terms responsible for such first-order effects. Consider an off-diagonal dissipator term as in Eq.~\eqref{Eq:OnePauliGates-OffDiagD}
\begin{align}
\mathcal{D}_\beta & = \beta_{jk} \mathcal{D}[P_j,P_k] = \beta_{jk} (P_j \bullet P_k^{\dag} - \frac{1}{2}\{P_k^{\dag}P_j, \bullet \}),\quad j\neq k.
\end{align}
 
To first order in perturbation, the contribution of $\beta_{jk}$ to the fidelity asymmetry is
\begin{align}
\Delta f^{(1)}_{mn,jk}
&= \frac{1}{D}\int_0^{\tau_g}\!dt'\;\beta_{jk}\,
\Tr\!\Big[P_m^\dagger\Big(P_j(t')\,P_m\,P_{k}^\dagger(t')
-\frac12\{P_{k}^\dagger(t')P_j(t'),P_m\}
\Big)\Big]\nonumber\\
&-\frac{1}{D}\int_0^{\tau_g}\!dt'\;\beta_{jk}\,
\Tr\!\Big[P_m^\dagger P_l^\dagger\Big(
P_j(t')\,P_l P_m\,P_{k}^\dagger(t')
-\frac12\{P_{k}^\dagger(t')P_j(t'),P_l P_m\}
\Big)\Big].
\label{Eq:MultiPauliGates-Delta f_mn1}
\end{align}
where we have set $s_{ab}^2=1$. The second (anticommutator) contributions cancel by the trace cyclicity, 
\begin{align}
\Tr\!\left[P_m^\dagger\{P_{k}^\dagger(t')P_j(t'),P_m\}\right]=\Tr\!\left[P_m^\dagger P_l^\dagger\{P_{k}^\dagger(t')P_j(t'),P_l P_m\}\right] \;,
\end{align}
so that we obtain the analogue of Eq.~\eqref{Eq:OnePauliGates-OffDiagD2} as:
\begin{align}
\Delta f_{mn,jk}^{(1)} = \frac{2}{D}\int_{0}^{\tau_g} dt' \beta_{jk} \Tr\{P_m^{\dag}\Pi_{P_l}^{-}[P_j(t')]P_m P_{k}^{\dag}(t')\}.
\label{Eq:MultiPauliGates-Delta f_mn2}
\end{align}

Since $P_l$ is a Pauli, from the commuting generator set, conjugation by it acts as a $\pm$ sign on each Pauli component of $P_j(t)$. Thus the bracket in~\eqref{Eq:MultiPauliGates-Delta f_mn2} projects onto the odd part of $P_j(t)$
that anticommutes with $P_l$ as:
\begin{align}
\Pi_{P_l}^{-}\!\big(P_j(t)\big) \equiv \frac{1}{2}[P_j(t)-P_l^\dagger P_j(t)P_l] =
\begin{cases}
P_j(t), & \{P_j,P_l\}=0,\\
0, & [P_j,P_l]=0.
\end{cases} \;,
\label{Eq:MultiPauliGates-Gate projector}
\end{align}
Similarly, given that $[P_m,P_l]=0$, we could have alternatively rewritten Eq.~\eqref{Eq:MultiPauliGates-Delta f_mn2} in terms of the projector of $P_k(t')$ as $\Tr[P_m^\dagger P_j(t') P_m (P_k(t')-P_l P_k(t')P_l^\dagger)]$. Therefore, altogether, the nonzero contributions in Eq.~\eqref{Eq:MultiPauliGates-Delta f_mn2} reduce to terms of the form $\Tr[P_m^\dagger P_j(t) P_m P_{k}^\dagger(t)]$, which arise only when $\{P_j,P_l\}=\{P_k,P_l\}=0$.

To get an idea of the general structure of non-zero elements we first rewrite Eq.~\eqref{Eq:MultiPauliGates-4termDecomp} in the compact form
\begin{align}
P_j(t) &= \sum_{r\in \{G,I\}} C_{jr}(t)\, P_r P_j , \\
P_{k}^\dagger(t) &= \sum_{r'\in \{G,I\}} C_{kr'}^{*}(t)\, P_k P_{r'} .
\end{align}
Conjugation of $P_j(t)$ by $P_m$ simply changes the sign of those terms that anticommute with $P_m$. Consequently,
\begin{align}
P_m^\dagger P_j(t) P_m
= \sum_{r \{G,I\}} s_{jmr}\, C_{jr}(t)\, P_r P_j ,
\end{align}
where the sign coefficient $s_{jmr}=\pm 1$ indicates whether $P_m$ commutes ($+1$) or anticommutes ($-1$) with the operator $P_r P_j$. Therefore we can rewrite the trace as
\begin{align}
\Tr[P_m^\dagger P_j(t) P_m P_{k}^\dagger(t)] = \sum\limits_{r,r'\in \{G,I\}}s_{jmr}C_{jr}(t)C_{kr'}^*(t) \Tr[P_rP_jP_kP_{r'}] \;,
\end{align}
which can \textit{in principle} be non-zero when
\begin{align}
P_j P_k \propto P_{r'}^{\dag} P_r \in \{I,P_a,P_b,P_c\} \;.
\end{align}

To identify the non-zero elements for the particular Pauli pair $(P_m, P_n)$, and show why the $P_jP_k\propto I$ condition cancels out, let us assume that $(P_m, P_n)$ are transformed by $P_c$, i.e. the $(1,1)$ subspace of Eq.~\eqref{Eq:MultiPauliGates-4termDecomp}. Having found that $\{P_j,P_l\}=\{P_k,P_l\}=0$ for $P_l=P_c$, we conclude that $P_j$ and $P_k$ belong to a different subspace compared to $P_l$, in this case either of the $(0,1)$ or the $(1,0)$ subspaces. This leads us to consider the following two cases:  

\textbf{Case (i):$(b_{ja},b_{jb})=(b_{ka},b_{kb})$} 

Consider first the case $(b_{ja},b_{jb})=(b_{ka},b_{kb})=(1,0)$. Moreover, we have $\{P_m,P_a\}=\{P_m,P_b\}=0$. According to the second line of Eq.~\eqref{Eq:MultiPauliGates-4termDecomp} we find:
\begin{align}
\begin{split}
\Tr[P_m^\dagger P_j(t) P_m P_{k}^\dagger(t)] & = \pm \Tr \Big\{[c^2 P_j - i\,c s\, P_a P_j + s_{ab}\, s^2 P_b P_j + i\,c s\, P_c P_j] [c^2 P_k - i\,c s\, P_k P_a - s_{ab}\, s^2 P_k P_b - i\,c s\, P_k P_c]\Big\} \\
& = \pm \Big\{(c^2-s^2) \Tr[P_jP_k] - 2ics\Tr[P_jP_kP_a] \Big\} \\
& = \pm \Big\{\cos(2\theta(t)) \Tr[P_jP_k] - i\sin(2\theta(t))\Tr[P_jP_kP_a]\Big\} \;,
\end{split}
\end{align}
where the overall $\pm$ sign comes from the commutation of $P_m$ and $P_j$. Note that $\int_{0}^{\tau_g}dt \cos(2\theta(t))=0$, and $\int_{0}^{\tau_g}dt \sin(2\theta(t))=1/\omega_g$, for the Clifford rotation $\omega_g \tau_g = \pi/2$. Therefore, only the prefactor of the $\Tr[P_jP_kP_a]$ term survives. 

By the same token, for the case $(b_{ja},b_{jb})=(b_{ka},b_{kb})=(0,1)$, one finds $\Tr[P_m^\dagger P_j(t) P_m P_{k}^\dagger(t)] = \pm \Big\{\cos(2\theta(t)) \Tr[P_jP_k] - i\sin(2\theta(t))\Tr[P_jP_kP_b]\Big\}$, leading to overall nonzero terms according to $\Tr[P_jP_kP_b]$.

\textbf{Case (ii):$(b_{ja},b_{jb})\neq(b_{ka},b_{kb})$}

Here, we have $P_j \neq P_k$. Consider first the case $(b_{ja},b_{jb})=(1,0)$ and $(b_{ka},b_{kb})=(0,1)$. Based on the second and third lines of Eq.~\eqref{Eq:MultiPauliGates-4termDecomp} we find:
\begin{align}
\begin{split}
\Tr[P_m^\dagger P_j(t) P_m P_{k}^\dagger(t)] & = \pm \Tr \Big\{[c^2 P_j - i\,c s\, P_a P_j + s_{ab}\, s^2 P_b P_j + i\,c s\, P_c P_j] [c^2 P_k - i\,c s\, P_k P_b - s_{ab}\, s^2 P_k P_a - i\,c s\, P_k P_c]\Big\} \\
& = \pm \Big\{c^2\Tr[P_jP_k] - ics\Tr[P_jP_kP_a]-ics\Tr[P_jP_kP_b]-s^2\Tr[P_jP_kP_c] \Big\} \;,
\end{split}
\end{align}
where here $\Tr[P_jP_k]=0$ since $P_j \neq P_k$. The integral of the prefactors for the rest of the terms are non-zero. A similar conclusion is reached for the case $(b_{ja},b_{jb})=(0,1)$ and $(b_{ka},b_{kb})=(1,0)$. 

Lastly, we note that the cases in which the Pauli pairs are transformed by $P_a$ or $P_b$ can be argued similarly, given the symmetric structure of Eq.~\eqref{Eq:MultiPauliGates-4termDecomp} with respect to $\{P_a,P_b,P_c\}$.  

The derivation so far assumed three commuting non-identity Pauli generators. We next comment on the case with just two commuting Paulis as in the case of iSWAP gate. Therefore, assuming the generator set $G\equiv \{P_a, P_b\}$, the interaction-frame Pauli operators are found as 
\begin{align}
P_j(t)=
\begin{cases}
P_j, & (b_{aj},b_{bj})=(0,0),\\[3pt]
c P_j + i\,s\, P_a P_j, & (b_{aj},b_{bj})=(1,0),\\[3pt]
c P_j + i\,s\, P_b P_j, & (b_{aj},b_{bj})=(0,1),\\[3pt]
c^2 P_j + i\,c s\,(P_a P_j+P_b P_j)-\, s^2\, P_aP_b P_j, & (b_{aj},b_{bj})=(1,1),
\end{cases}
\end{align}
Here, the algebra is distinct, where for those Pauli pairs transformed by $P_l \in G=\{P_a,P_b\}$ we have $\{P_l, P_m\}=\{P_l, P_n\}=0$, while for pairs transformed by the product $P_aP_b \propto P_c$, even though not part of the initial generator, we have $[P_l, P_m]=[P_l, P_n]=0$. It can be shown that the nonzero coefficients obey $P_jP_k\propto P_l \in G$.     

To summarize, the set of all dissipator elements that cause first-order fidelity asymmetry obey Result~\ref{result:MultiPauli} of the main text:
\begin{align}
\Delta f^{(1)}_{mn,jk}\neq 0
\quad\Longleftrightarrow\quad
P_jP_k\propto P_l\ \ \text{and}\ \ \{P_j,P_l\}=\{P_k,P_l\}=0 \;,
\label{eq:CZ_selection_rule}
\end{align}
for any $P_l\in G$.
%%%%%%%%%%%%%%%%%%%%%%%%%%%%%%%%%%%%%%%%%%%%%%%%%

% Appendix: %%%%%%%%%%%%%%%%%%%%%%%%%%%%%
\section{Gates generated by a primary Pauli plus fractional commuting Pauli terms}
\label{App:MultiPauliWithPhase1}

In this Appendix, we analyze gates that are physically implemented using a primary Pauli generator with additional commuting Pauli terms that accumulate fractional rotation angles during the gate but are coherently corrected at its completion. Following the same methodology used in Appendix~\ref{App:MultiPauliGates}, we derive Result~\ref{result:MultiPauliPhase1}. Two key differences arise in comparison to Result~\ref{result:MultiPauli} and Appendix~\ref{App:MultiPauliGates}:
(i) the relevant anticommutation selection rules depend solely on the single primary generator Pauli; and
(ii) products of the distinct partial angles lead to non‑vanishing contributions from the diagonal elements of the dissipator.

The physical gate Hamiltonian is taken as
\begin{align}
H_g
= \frac{\omega_g}{2}\!\left(P_g + \sum_{l\in S}\alpha_l\,P_l\right) \;,
\label{Eq:MultiPauliWithPhase1-description}
\end{align}
where all generator Paulis commute $[P_g,P_l]=[P_l,P_l']=0, \quad \forall l,l' \in S$. Following Appendix~\ref{App:MultiPauliGates}, we denote the commuting generators $\{P_a,P_b,P_c\}$ with $P_a\equiv P_g$ and $P_c=s_{ab}P_aP_b$ ($s_{ab}\in\{\pm1\}$). Define angles $\theta_a(t)\equiv \omega_g t$, $\theta_l(t)\equiv \alpha_l \omega_g t$ for $l\in\{b,c\}$, and abbreviations $c_x(t)\equiv \cos\theta_x(t)$, $s_x(t)\equiv \sin\theta_x(t)$ for $x\in\{a,b,c\}$. 

Then, the interaction-frame Pauli operators are obtained similar to Eq~(\ref{Eq:MultiPauliGates-4termDecomp}), but with unequal angles, as
\begin{align}
P_j(t)=
\begin{cases}
P_j, & (b_{aj},b_{bj})=(0,0),\\[4pt]
c_a c_c\, P_j
+ i\,s_a c_c\, P_a P_j
- s_{ab}\, s_a s_c\, P_b P_j
+ i\,c_a s_c\, P_c P_j,
& (b_{aj},b_{bj})=(1,0),\\[8pt]
c_b c_c\, P_j
+ i\,s_b c_c\, P_b P_j
- s_{ab}\, s_b s_c\, P_a P_j
+ i\,c_b s_c\, P_c P_j,
& (b_{aj},b_{bj})=(0,1),\\[8pt]
c_a c_b\, P_j
+ i\,s_a c_b\, P_a P_j
+ i\,c_a s_b\, P_b P_j
- s_{ab}\, s_a s_b\, P_c P_j,
& (b_{aj},b_{bj})=(1,1),
\end{cases}
\label{Eq:MultiPauliWithPhase1-ExplicitPjI}
\end{align}
Equivalently, with $\mathcal A=\langle P_a,P_b,P_c\rangle=\{I,P_a,P_b,P_c\}$, $P_j(t)$ can be compactly expressed
\begin{align}
P_j(t)=\sum_{r\in\mathcal{A}}C_{jr}(t)P_rP_j \;,
\label{Eq:MultiPauliWithPhase1-CompatctPjI(t)}
\end{align}
with coefficients $C_{jr}(t)$ given in Eq.~\eqref{Eq:MultiPauliWithPhase1-ExplicitPjI}.

For the Clifford pair $(P_m,P_n)$ with $P_n=U_g^\dagger\,P_m\,U_g=iP_g P_m$,
the anti‑commutator terms cancel as in Appendices~B–C, giving
\begin{align}
\begin{split}
\Delta f^{(1)}_{mn,jk} & = \frac{1}{D}\int_0^{\tau_g}\!dt'\;\beta_{jk}\, \Tr\!\Big[P_m^\dagger\big(P_j(t')-P_g^\dagger P_j(t')P_g\big)\,P_m\,P_{k}^\dagger(t')\Big]\\
&=\frac{2}{D}\!\int_0^{\tau_g}\!dt'\;\beta_{jk}\,
\Tr\!\Big[P_m^\dagger\,\Pi_{P_g}^-\!\big(P_j(t')\big)\,P_m\,P_{k}^\dagger(t')\Big] \;,
\label{Eq:MultiPauliWithPhase1-Deltaf-compact}
\end{split}
\end{align}
where $\Pi_{P_g}^-(X)\equiv\tfrac{1}{2}(X-P_g^\dagger X P_g)$ projects onto the part \emph{odd} under $P_g$. Since $[P_r,P_g]=0$ for all $P_r\in\mathcal{A}$, one finds
\begin{align}
\Pi_{P_g}^-\!\big(P_j(t)\big)=
\begin{cases}
P_j(t), & \{P_j,P_g\}=0,\\
0, & [P_j,P_g]=0.
\end{cases}
\end{align}
Thus \eqref{Eq:MultiPauliWithPhase1-Deltaf-compact} is nonzero only if $\{P_j,P_g\}=0$. Following the same argument as under Eq.~\eqref{Eq:MultiPauliGates-Gate projector}, re-expressing the projector in terms of $P_k(t')$, one concludes that $\{P_k,P_g\}=0$. 

Inserting expansion~(\ref{Eq:MultiPauliWithPhase1-CompatctPjI(t)}), and similarly for $P_k(t)=\sum_{r'\in\mathcal{A}} C_{kr'}(t)\,P_{r'}P_k$, into Eq.~\ref{Eq:MultiPauliWithPhase1-Deltaf-compact} we find:
\begin{align}
\Delta f^{(1)}_{mn,jk}
=\sum_{r,r'\in\mathcal A}\frac{2}{D}\!\int_0^{\tau_g}\!dt'\;\beta_{jk}
C_{jr}(t')\,C_{kr'}^*(t')\;
\Tr\!\big[P_m^\dagger\,(P_rP_j)\,P_m\,(P_{r'}P_k)\big].
\label{Eq:MultiPauliWithPhase1-Sum}
\end{align}
In contrast to the common-angle gates discussed in Appendix~\ref{App:MultiPauliGates}, the time integrals $\int_{0}^{\tau_g}dt' C_{jr}(t')C^*_{kr'}(t')$ that involve multiple fractional rotation angles do not vanish in the present setting. Consequently, the non‑zero contributions are determined solely by Pauli orthogonality: the trace is non‑zero if a Pauli component matches:
\begin{align}
P_r P_j=\pm P_{r'}P_k
\Longleftrightarrow
P_j P_k = P_{r'}^{-1}P_{r}\in \mathcal{A} \;.
\label{Eq:MultiPauliWithPhase1-Trace condition}
\end{align}

In summary, non-zero first-order asymmetries are caused by the dissipative noise elements described by the following rule:
\begin{boxedresultdef}
\label{result:MultiPauliPhase1}
Consider a Clifford gate which is \textit{physically} generated by $H_g=\frac{\omega_g}{2}\!\left(P_g+\sum_{l\in S}\alpha_l P_l\right)$ with $[P_g,P_l]=[P_l,P_{l'}]=0$ for all $l,l'$, and let $\mathcal{A}=\langle P_g,\{P_l\}_{l\in S}\rangle$ be the Abelian subgroup they generate (including identity). Assume the target Clifford gate is the rotation by $P_g$, while the accompanying $P_l$-generated phases are perfectly corrected via end-of-gate corrections $e^{i(\phi_l/2) P_l}$ with $\phi_l=\alpha_l\omega_g\tau_g$. For arbitrary coherent and dissipative noise as in Eqs.~(\ref{eq:LindEq})--(\ref{eq:Def_of_Dbeta}),
the first-order Pauli-fidelity asymmetry comes from dissipator elements $\beta_{jk}$ satisfying
\begin{align}
P_jP_k\in\mathcal{A}, \quad \{P_j,P_g\}=\{P_k,P_g\}=0 \;.
\end{align}
\end{boxedresultdef}
We note that Result~\ref{result:MultiPauliPhase1} assumes nonzero fractional angles for \textit{all} commuting terms, hence gives the largest set of allowed dissipator elements that can cause asymmetry. For two qubits, $|\mathcal A|=4$ generically, giving $8|\mathcal A|=32$ ordered elements (including diagonals). When one of the fractional phases is set to zero, several first‑order $\bm{\beta}$ contributions cancel, leaving only 20 elements: 8 with $P_jP_k \propto P_g$ and 4 in each of the remaining commuting‑generator sectors.
%%%%%%%%%%%%%%%%%%%%%%%%%%%%%%%%%%%%%%%%%%%%%%%%%%%

%%%%%%%%%%%%% Table: asymmetry rules %%%%%%%
\begin{table*}[t]
\centering
\begin{tabular}{|C{0.17\textwidth}|C{0.25\textwidth}|C{0.05\textwidth}|C{0.30\textwidth}|C{0.05\textwidth}|C{0.05\textwidth}|}
\hline
\textbf{Gate family} &
\textbf{Gate instance} &
\textbf{$\mathcal{H}$ asym.} &
\textbf{$\mathcal{D}$ asym.} &
\textbf{$T_{1\downarrow}$ asym.} &
\textbf{$T_{2\phi}$ asym.} \\
\hline\hline

(1) Single-Pauli generator (R1) &
\makecell{$ZZ_{\pi/2}$} &
$O(\delta^2)$ &
\shortstack[c]{$O(\beta_{jk}):$ \\ $P_jP_k\propto P_g, \; \{P_j,P_g\}=\{P_k,P_g\}=0$} &
$O(\beta_{\downarrow}^2)$ &
$O(\beta_{\phi}^2)$ \\
\hline

(2) Common-angle commuting generators (R2) &
\shortstack[c]{%
$\mathrm{CZ}$,
$\mathrm{CNOT}$,
$\mathrm{iSWAP}$,
$\mathrm{SWAP}$
} &
$O(\delta^2)$ &
\shortstack[c]{$O(\beta_{jk}):$ \\ $ P_jP_k\propto P_l, \{P_j,P_l\}=\{P_k,P_l\}=0$ \\ $P_l\in G\!\setminus\!\{I\}$} &
$O(\beta_{\downarrow}^2)$ &
$O(\beta_{\phi}^2)$ \\
\hline

(3) Commuting generators corrected into a common-angle gate (R3) &
\shortstack[c]{%
$\mathrm{CZ}$ synthesized from dominant \\ $ZZ$ with $IZ, ZI$ \\
phase compensation
} &
$O(\delta^2)$ &
\shortstack[c]{$O(\beta_{jk}):$ \\ $ P_jP_k\propto P_l,\ \{P_j,P_l\}=\{P_k,P_l\}=0 $\\ $P_jP_k\propto I,\ \{P_j,P_{l'}\}=\{P_k,P_{l'}\}=0 $ \\
$ P_l,P_{l'} \in \mathcal{A}\!\setminus\!\{I\}$} &
$O(\beta_{\downarrow}^2)$ &
$O(\beta_{\phi}^2)$ \\
\hline

(4) Primary-generator + phase removal (R4) &
\shortstack[c]{%
$ZZ_{\pi/2}$ with residual $IZ, ZI$ \\ phases
removed by \\ end-of-gate corrections
} &
$O(\delta^2)$ &
\shortstack[c]{%
$O(\beta_{jk}):$ \\ $ P_jP_k\in\mathcal{A}, \{P_j,P_g\}=\{P_k,P_g\}=0$\\
$\mathcal{A}=\langle P_g,\{P_l\}\rangle$
} &
$O(\beta_{\downarrow}^2)$ &
$O(\beta_{\phi}^2)$ \\
\hline
\end{tabular}
\caption{Summary of leading-order Pauli-fidelity asymmetry mechanisms across Clifford gate families as in Results~\ref{result:OnePauli}--\ref{result:MultiPauliPhase1}.}

\label{Tab:selection_rules}
\end{table*}
%%%%%%%%%%%%%%%%%%%%%%%%%%%%%%%%%%%%%%%%%%%

%%%%%%%%%%%% Appendix: Gates generated by multiple commuting Pauli terms corrected into a common-angle form %%%%%%%%%%%%%
\twocolumngrid
\section{Gates generated by multiple commuting Pauli terms corrected into a common-angle form}
\label{App:MultiPauliWithPhase2}
In this Appendix, we study gates implemented via a primary Pauli generator supplemented by commuting Pauli terms that accrue fractional rotations and are coherently corrected to a common‑angle form. This case lies conceptually between Appendices~\ref{App:MultiPauliGates} and~\ref{App:MultiPauliWithPhase1}: The \emph{physical} gate evolution is generated by a sum of commuting Pauli terms with unequal rotation angles, as in Appendix~\ref{App:MultiPauliWithPhase1}, while the \emph{Pauli pairing} relevant for the asymmetry corresponds to a compensated \emph{common-angle} Clifford target gate, as in Appendix~\ref{App:MultiPauliGates}. Consequently, the interaction-frame operators \(P_j(t)\) and \(P_k(t)\) follow Eq.~\eqref{Eq:MultiPauliWithPhase1-ExplicitPjI}, whereas the odd projector is determined by the target-gate pairing, as in Appendix~\ref{App:MultiPauliGates}.

For a Pauli pair $(P_m,P_n)$ transformed by the action of the common-angle Clifford it follows similar to Eq.~\eqref{Eq:MultiPauliGates-CliffordPairing} that $P_n = U_g^\dagger\,P_m\,U_g = -s_{ab}\,P_l\,P_m$ with Pauli $P_l\in G$. Then, the first-order contribution of a single dissipator element $\beta_{jk}$ to the Pauli-fidelity asymmetry can be written as
\begin{align}
\Delta f^{(1)}_{mn,jk}
= \frac{2}{D}\int_{0}^{\tau_g}\!dt\;\beta_{jk}\,
\mathrm{Tr}\!\left[P_m^\dagger\,
\Pi^{-}_{P_l}\!\big(P_j(t)\big)\,
P_m\, P_{k}^\dagger(t)\right] \;.
\label{Eq:MultiPauliPhase2-universal}
\end{align}
Conjugation by $P_l$ acts as a $\pm$ sign on each Pauli component, hence the projector in Eq.~\eqref{Eq:MultiPauliPhase2-universal} enforces $\{P_j,P_l\}=0$,
and, equivalently by moving the odd projector onto $k$ inside the trace (as in Appendix~\ref{App:MultiPauliGates}), we find the aggregate condition
\begin{align}
\Delta f^{(1)}_{mn,jk}\neq 0
\quad\Longrightarrow\quad
\{P_k,P_l\}=\{P_j,P_l\}=0 \;.
\label{Eq:MultiPauliPhase2-oddj}
\end{align}

Under the setting of Appendix~\ref{App:MultiPauliWithPhase1}, with unequal-angle commuting generators, the trace in Eq.~\eqref{Eq:MultiPauliPhase2-universal} can be nonzero only when Pauli orthogonality permits a matching of Pauli components. This yields the same product constraint as in Appendix D:
\begin{align}
\Delta f^{(1)}_{mn,jk}\neq 0
\quad\Longrightarrow\quad
P_jP_k \in \mathcal{A} \;.
\label{Eq:MultiPauliPhase2-productA}
\end{align}
where $\mathcal{A}$ is the Abelian generator subgroup. Combining Eq.~\eqref{Eq:MultiPauliPhase2-productA} with the pair-dependent odd constraints
Eq.~\eqref{Eq:MultiPauliPhase2-oddj} gives two distinct surviving cases, which correspond exactly to the two lines stated in Result~\ref{result:MultiPauliPhase2}.

\paragraph*{Case 1: generator sectors ($P_jP_k\propto P_l$).} This is the direct analogue of the component-wise selection rule of Appendix~\ref{App:MultiPauliGates}, now evaluated with interaction-frame operators generated by the unequal-angle physical evolution (Appendix~\ref{App:MultiPauliWithPhase1}).
The derivation here follows the same steps as in Appendix~\ref{App:MultiPauliGates}, resulting in the selection rule
\begin{align}
P_jP_k \propto P_l, \quad \{P_j,P_l\}=\{P_k,P_l\}=0, \quad P_l\in\mathcal{A}\setminus\{I\} \;.
\label{Eq:MultiPauliPhase2-case1}
\end{align}

\paragraph*{Case 2: Identity sector ($P_jP_k\propto I$).}
When the orthogonality condition selects the identity sector, $P_jP_k \propto I$, the product constraint \emph{does not} specify a unique non-identity generator: $P_jP_k\propto I$ implies $P_k=\pm P_j$ (up to phase), so Pauli orthogonality alone cannot distinguish which $P_l\in\mathcal{A}\setminus\{I\}$ is responsible for a target-gate pairing. Consequently, the surviving first-order contribution in this sector is controlled entirely by the odd projector, i.e., by whether the diagonal component is odd under the \emph{pair-dependent} generator. For a fixed conjugate pair $(P_m,P_n)$ (fixed $P_l$), one requires
\begin{align}
P_jP_k \propto I,
\quad
\{P_j,P_l\}=\{P_k,P_l\}=0 \;.
\label{Eq:MultiPauliPhase2-case2}
\end{align}
Taking the union over all target-gate pair generators $P_l\in\mathcal{A}\setminus\{I\}$ yields the identity-sector condition stated in Result~\ref{result:MultiPauliPhase2}. This explains why the $P_jP_k\propto I$ sector is ``larger'': unlike Case~1, the identity product does not pick out a single generator sector, so any non-identity generator appearing as a target-gate pairing can activate the odd projector. 

In summary, Eqs.~\eqref{Eq:MultiPauliPhase2-case1} and~\eqref{Eq:MultiPauliPhase2-case2} are the two selection rules quoted in Result~\ref{result:MultiPauliPhase2}.  We conclude our analysis by summarizing, in Table~\ref{Tab:selection_rules}, the Pauli noise asymmetry scaling and the associated selection rules governing first‑order dissipative asymmetry for the considered family of gates.
%%%%%%%%%%%%%%%%%%%%%%%%%%%%%%%%%%%%%%%%%%%%%%%%%%%%%%%%%%%%

%%%%%%%%%%%% Appendix: Numerical sims %%%%%%%%%%%%%%%%%
\section{Simulation of Pauli noise symmetry from Lindbladian dynamics}
\label{App:AsymmSim}

%%%Fig: dissipator asymmetry patern 2 %%%%
\begin{figure}[t!]
\centering
\includegraphics[scale=0.271]{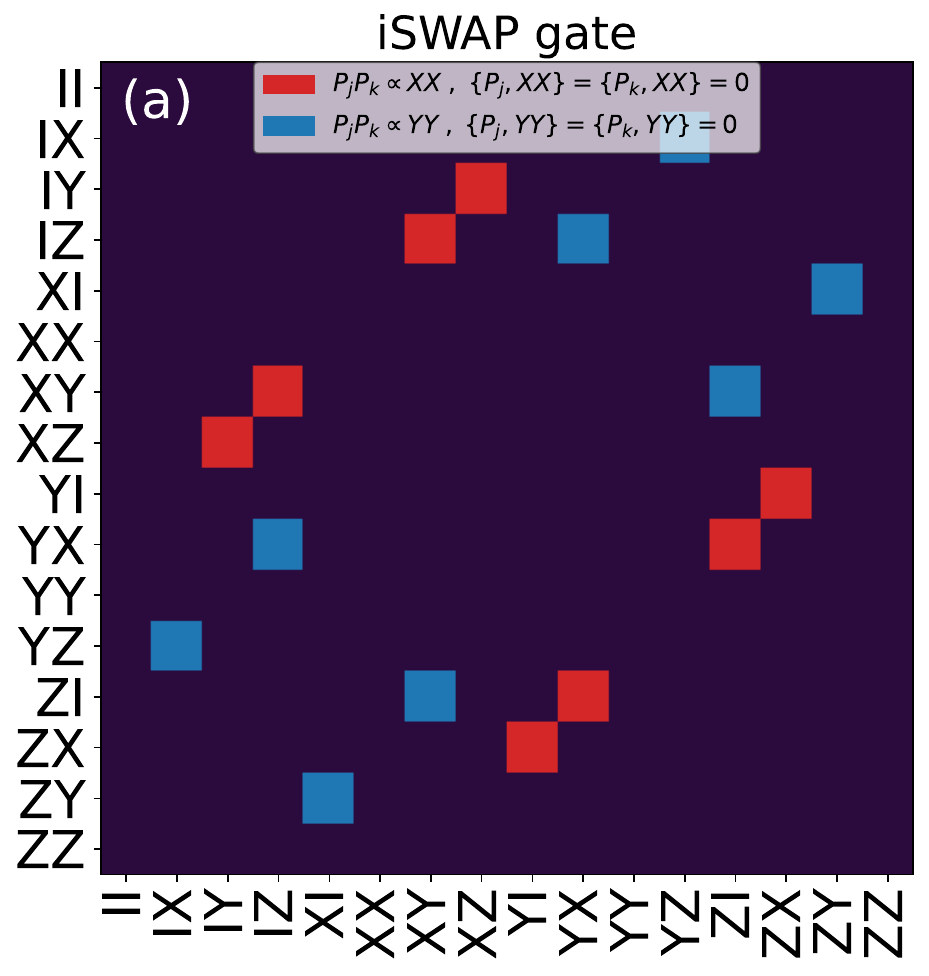}
\includegraphics[scale=0.271]{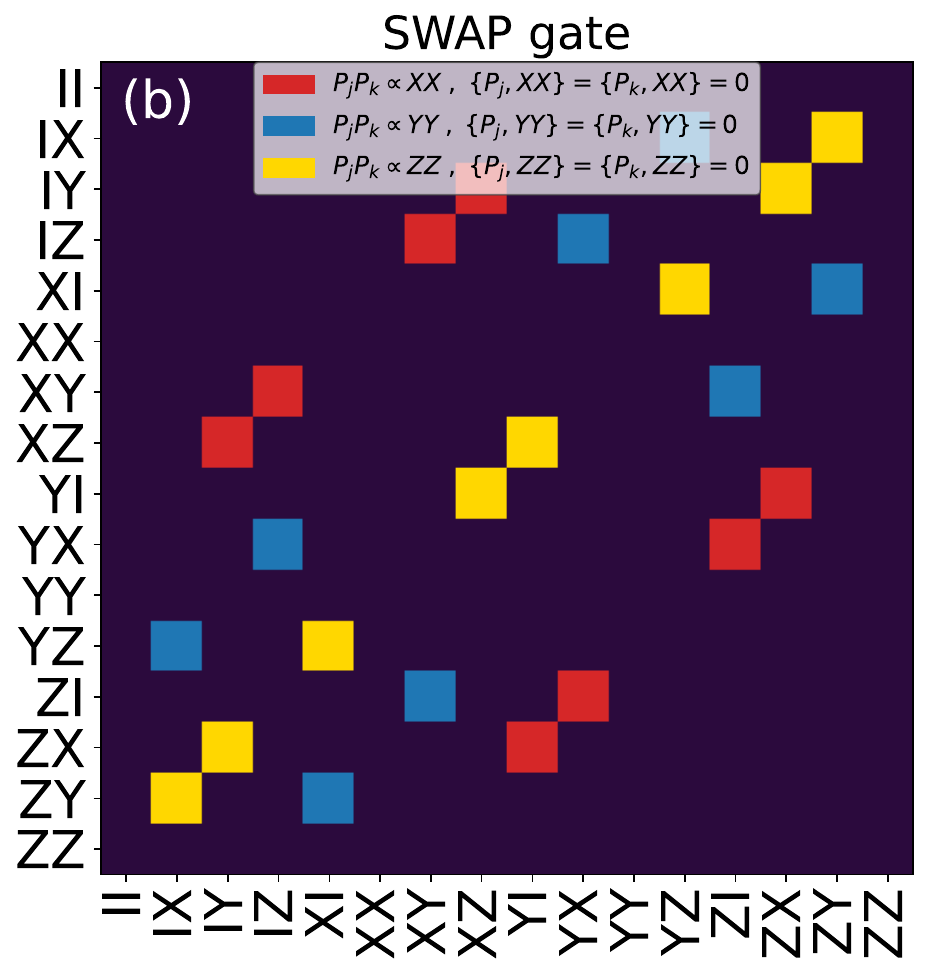}
\includegraphics[scale=0.271]{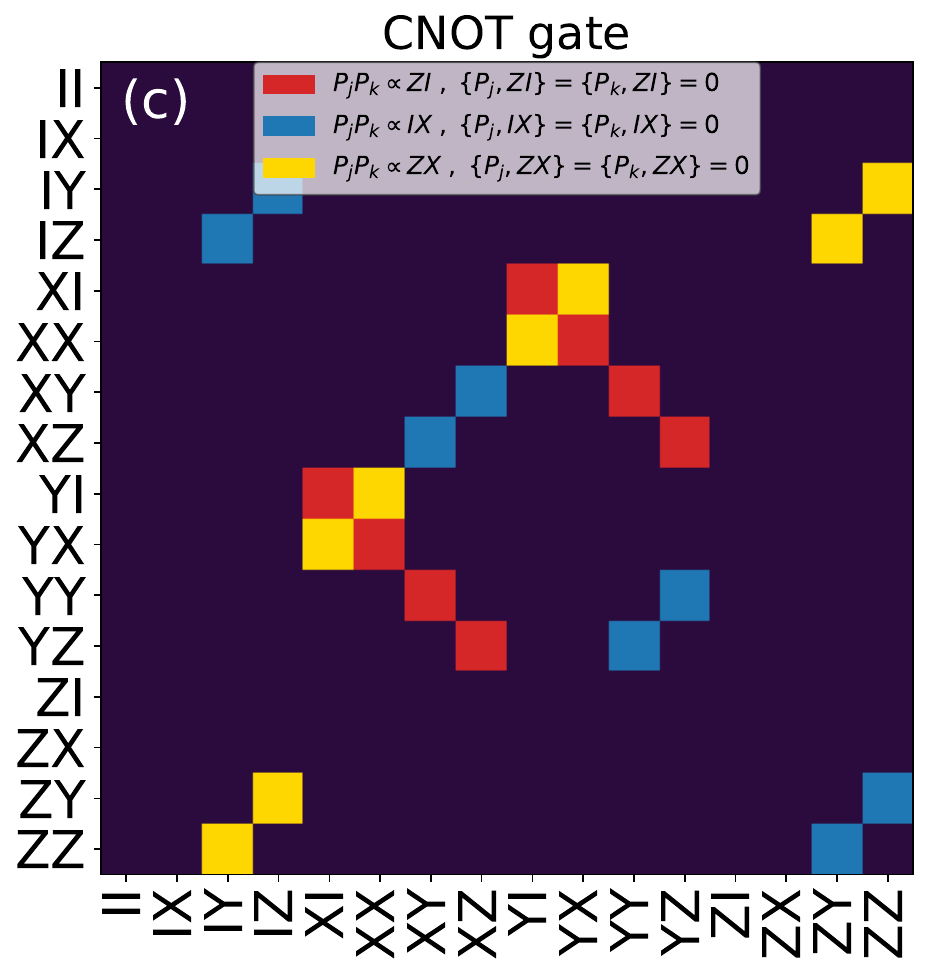}
\caption{\textbf{Leading-order pattern of Pauli-fidelity asymmetry induced by dissipative noise}--- Additional notable gates belonging to the case of Result~\ref{result:MultiPauli}: (a) iSWAP, (b) SWAP, (c) CNOT. For gates whose generators decompose into commuting Pauli terms, contributions are grouped by generator component.}    
\label{fig:dissipator_asymmetry_patterns2}
\end{figure}
%%%%%%%%%%%%%%%%%%%%%%%%%%

In this Appendix, we further validate our analytical characterization of Pauli noise symmetry in Appendices~\ref{App:LindPert}--\ref{App:MultiPauliWithPhase2} through numerical simulations. We simulate the Lindblad noise model~\eqref{eq:LindEq}--\eqref{eq:Def_of_Dbeta}, assuming time-independent ideal gate and noise parameters, and compute the noise channels via direct exponentiation following Eqs.~\eqref{eq:Def_of_Lambda^G}--\eqref{eq:Def_of_f}.

We quantify noise asymmetry in the PTM representation by comparing the original noise channel with its gate-conjugated counterpart, defining
\begin{align}
\Delta \Lambda^{\mathcal{G}_{\text{t}}} \equiv \mathcal{U}_g\Lambda^{\mathcal{G}_{\text{t}}} \mathcal{U}_g^{-1} - \Lambda^{\mathcal{G}_{\text{t}}} \;.
\label{Eq:AsymmSim-Def_of_DelLamba^G}
\end{align}
The diagonal elements of the PTM in Eq.~\eqref{Eq:AsymmSim-Def_of_DelLamba^G} correspond to the fidelity asymmetry $\Delta f$. To characterize the order of Pauli noise asymmetry, we vary the strength of each individual Hamiltonian and dissipative term and fit the resulting $\Delta f$ as a function of the corresponding scaling factor. 

%%%Fig: psd dissipator asymmetry patern %%%%
\begin{figure}[t!]
\centering
\includegraphics[scale=0.205]{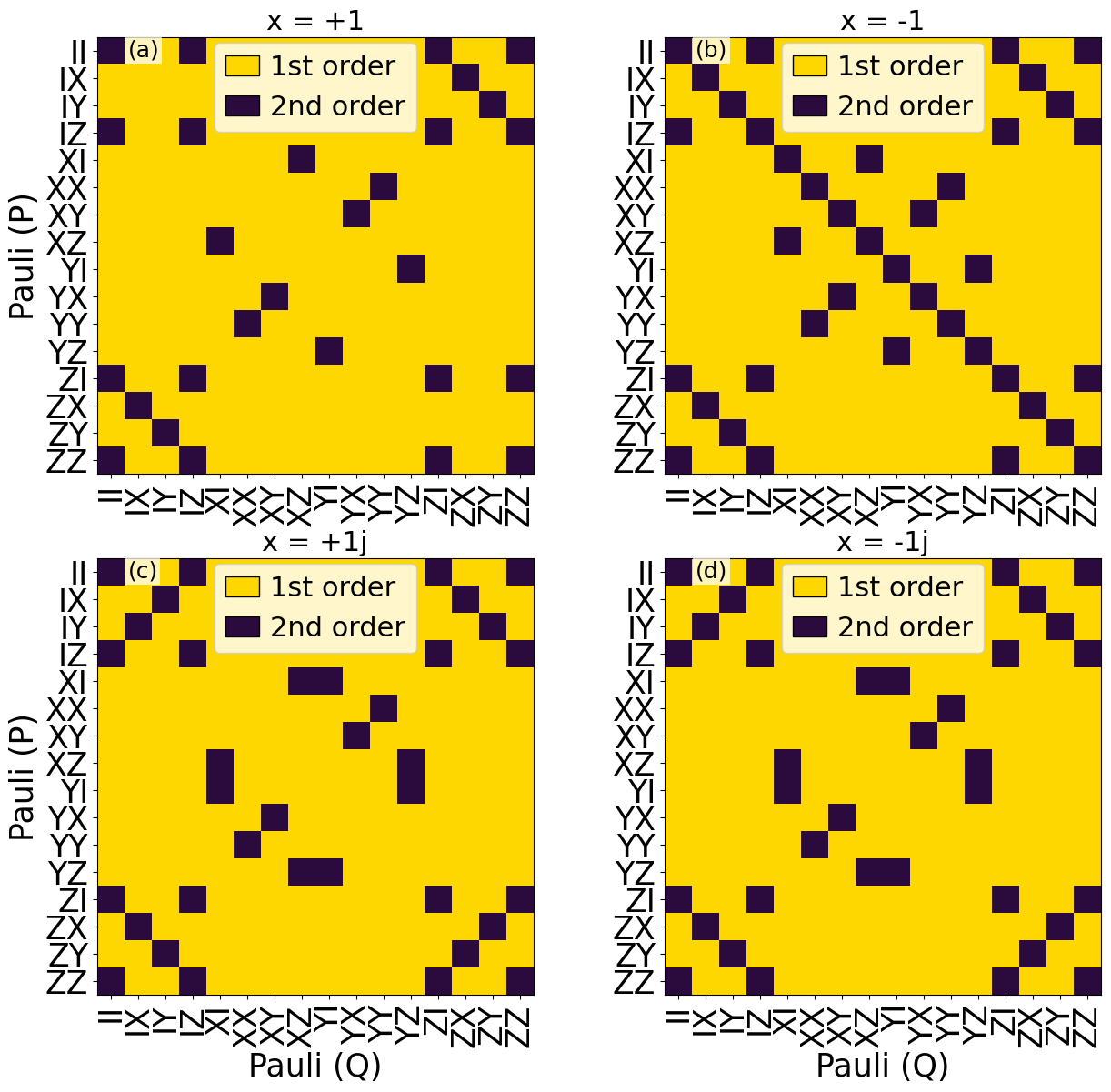}
\caption{\textbf{Pattern of Pauli-fidelity asymmetry induced by PSD dissipative noise for CZ gate with phase compensation}---Panels (a)--(d) represent the four possible linear combinations of Paulis, required for a PSD $\bm{\beta}$ matrix, as $\mathcal{D}[P+Q]$, $\mathcal{D}[P-Q]$, $\mathcal{D}[P+iQ]$, and $\mathcal{D}[P-iQ]$, respectively. Yellow and blue pixels denote first- and second-order dependence of noise asymmetry~\eqref{Eq:AsymmSim-Def_of_DelLamba^G} on dissipative noise $\beta$. Standard $T_1$ relaxation/excitation and $T_{2\phi}$ pure dephasing lead to a second-order asymmetry. Here, the starting single-qubit $Z$ angles were set to $\alpha_{IZ}=0.1$ and $\alpha_{ZI}=0.2$ before compensation.}   
\label{fig:psd_dissipator_asymmetry_patterns}
\end{figure}
%%%%%%%%%%%%%%%%%%%%%%%%%%

We first corroborate the dissipative asymmetry patterns found for $ZZ_{\pi/2}$, CZ, and CZ with $IZ/ZI$ phase compensation shown in Fig.~\ref{fig:dissipator_asymmetry_patterns}. We further analyze other standard two-qubit Clifford gates belonging to case (2a) (Result~\ref{result:MultiPauli}) of the main text. Fig.~\ref{fig:dissipator_asymmetry_patterns2} shows the first-order fidelity asymmetry due to dissipative noise for iSWAP, SWAP and CNOT gates. Here, we numerically derive the patterns and verify consistency with Result~\ref{result:MultiPauli} by grouping contributions according to individual gate generators. For CNOT, the leading-order asymmetry is governed by generators $\{ZI, IX, ZX\}$. For iSWAP and SWAP, it is set by $\{XX, YY\}$ and $\{XX, YY, ZZ\}$, respectively.

We also numerically validate the selection rules while enforcing the physical constraint that the dissipator matrix $\beta$ is positive semidefinite (PSD). In particular, a single off-diagonal element $\beta_{jk}$ cannot be tuned independently without violating PSD; instead, a PSD rank-one dissipator is realized by a single Lindblad collapse operator $C\propto P+xQ$ with $x\in\{+1,-1,+i,-i\}$. Fig.~\ref{fig:psd_dissipator_asymmetry_patterns} shows the resulting first-order asymmetry patterns for the CZ gate with phase compensation (as in our experimental realization) obtained by sweeping over Pauli pairs $(P,Q)$ and constructing dissipators of the form $\mathcal{D}[P+xQ]$, corresponding to the four choices $x=\pm1,\pm i$ (panels (a)–(d)). For each such PSD dissipator we scale the overall noise strength and fit the induced diagonal entries of $\Delta\Lambda^{\mathcal{G}_{\text{t}}}$ [Eq.~\eqref{Eq:AsymmSim-Def_of_DelLamba^G}] to determine whether the associated Pauli-fidelity asymmetry $\Delta f$ is linear (first order) or quadratic (second order) in the dissipative rate $\beta$. The support of first-order contributions matches the analytical selection rule in Appendix ~\ref{App:MultiPauliWithPhase2} and Result~\ref{result:MultiPauliPhase2}. However, after accounting for the PSD mixing enforced by $C\propto P+xQ$, the conditions are more stringent, leading to dense first-order contributions (yellow background). Nevertheless, importantly, dissipators corresponding to standard single-qubit $T_1$ relaxation/excitation only induce Pauli-fidelity asymmetry at second order [see $IX,IY$ and $XI,YI$ pairs in panel (c)]. Moreover, standard $T_{2\phi}$ pure dephasing also leads to second-order asymmetry.

%%%%%%%%%%%%%%%%%%%%%%%%%%%%%%%%%%%%%%%%%%%%%%%%%%%%%%%

%%%%%%%%%%%% Appendix: Symmetric noise learning %%%%%%%%%%%%
\section{Pauli noise learning in the symmetric gauge}
\label{App:SymmetricGauge}

Here, we review how the symmetric-gauge condition can be used to fix the depolarizing gauge left undetermined by self-consistent Pauli-noise learning, thereby separating state-preparation and measurement contributions to SPAM. The construction follows the gauge framework of Ref.~\cite{chen2026efficient}, but adds the physically motivated Pauli-fidelity symmetry constraints derived in Results~\ref{result:OnePauli}--\ref{result:MultiPauliPhase2}.

In the self-consistent learning framework of Ref.~\cite{chen2026efficient}, experiments are constructed as state preparation $S$, a sequence of Clifford gates $\mathcal{G}$, followed by measurement of a Pauli observable $M$. For each noisy component $\mu\in\{S,M,\mathcal G\}$, let $f_P^\mu$ denote the Pauli fidelity associated with Pauli operator $P$. As in the main text, taking logarithms turns products of fidelities into a linear system. We denote log fidelities as
\begin{align}
x_P^\mu := -\log f_P^\mu \;,
\label{Eq:SymmetricGauge-x-def}
\end{align}
so that the measured quantities vector $\mathbf{y}$ depend linearly on log fidelities as
\begin{align}
\bm y = F\bm x \;.
\label{Eq:SymmetricGauge-linear-system}
\end{align}    
where $F$ is a design matrix set by the circuit structure and further locality assumptions.  

For depth-0 and depth-1 experiments of Fig.~\ref{fig:gauge_consistent_learning}, the relevant observables may be written as
\begin{align}
y^{(0)}_P &= x^S_P + x^M_P \;,
\label{Eq:SymmetricGauge-depth0}
\\
y^{(1)}_{P} &= x^S_P + x_P^{\mathcal{G}_{\text{t}}} + x^M_{U_g P U_g^\dagger} \;.
\label{Eq:SymmetricGauge-depth1}
\end{align}
Equation~\eqref{Eq:SymmetricGauge-depth0} shows that depth-0 data determine only the SPAM product, not the individual state-preparation and measurement terms.

The gauge action is induced by insertion of a generalized depolarizing channel $\mathcal D_{\bm\eta}$, under which
\begin{align}
\Lambda^S &\mapsto \mathcal D_{\bm\eta} \Lambda^S, \nonumber\\
\Lambda^M &\mapsto \Lambda^M \mathcal D_{\bm\eta}^{-1}, \nonumber\\
\Lambda^{\mathcal{G}_{\text{t}}}
&\mapsto
\mathcal D'_{\bm\eta} \Lambda^{\mathcal{G}_{\text{t}}} \mathcal D_{\bm\eta}^{-1} \;,
\label{Eq:SymmetricGauge-gauge-transform}
\end{align}
where $\mathcal D'_{\bm\eta}
\equiv \mathcal{U}_g^{-1} \mathcal D_{\bm\eta} \mathcal{U}_g$ is the gate-conjugated depolarization channel. If $d_P^{\bm\eta}$ is the Pauli fidelity of $\mathcal D_{\bm\eta}$ on $P$, and
\begin{align}
\gamma_P^{\bm\eta} := -\log d_P^{\bm\eta} \;,
\label{Eq:SymmetricGauge-gamma-def}
\end{align}
then the log-fidelity parameters transform as
\begin{align}
x^{S,\bm\eta}_P & =  x^S_P + \gamma_P^{\bm\eta} \;, 
\label{Eq:SymmetricGauge-gauge-s}\\
x^{M,\bm\eta}_P & = x^M_P - \gamma_P^{\bm\eta} \;, 
\label{Eq:SymmetricGauge-gauge-m}\\
x^{\mathcal{G}_{\text{t}},\bm\eta}_P &= x_P^{\mathcal{G}_{\text{t}}} + \gamma_{U_g P U_g^\dagger}^{{\bm\eta}} -\gamma_P^{\bm\eta} \;.
\label{Eq:SymmetricGauge-gauge-g}
\end{align}
These shifts leave Eqs.~\eqref{Eq:SymmetricGauge-depth0} and \eqref{Eq:SymmetricGauge-depth1} invariant, so the learned model is only defined up to this gauge freedom.

Under the quasi-local ansatz, the gauge reduces to $n$ single-qubit depolarizing parameters. Equivalently,
\begin{align}
\gamma_P^{\bm\eta}
=
\sum_{q=1}^n \eta_q \chi_q(P),
\qquad
\chi_q(P)=
\begin{cases}
1, & P_q\neq I\\
0, & P_q=I
\end{cases},
\label{Eq:SymmetricGauge-local-gamma}
\end{align}
where $\chi_q(P)$ is the Pauli component of $P$ acting on qubit $q$. Thus the entire gauge is specified by $\bm\eta=(\eta_1,\dots,\eta_n)$.

To fix $\bm\eta$, we impose the symmetric-gauge condition
\begin{align}
f_P^{\mathcal{G}_{\text{t}},\eta}
=
f_{U_g P U_g^\dagger}^{\mathcal{G}_{\text{t}},\eta}
\qquad
\text{for all conjugate pairs } (P,U_g P U_g^\dagger).
\label{Eq:SymmetricGauge-symmetric-condition}
\end{align}
In log form this becomes
\begin{align}
x_P^{\mathcal{G}_{\text{t}},\eta}
=
x_{U_g P U_g^\dagger}^{\mathcal{G}_{\text{t}},\eta} \;.
\label{Eq:SymmetricGauge-symmetric-log}
\end{align}
Substituting Eq.~\eqref{Eq:SymmetricGauge-gauge-g} on both sides gives
\begin{align}
2\Big(\gamma_{U_g P U_g^\dagger}^{\bm\eta}-\gamma_P^{\bm\eta}
\Big)= x_{U_g P U_g^\dagger}^{\mathcal{G}_{\text{t}}}-x_P^{\mathcal{G}_{\text{t}}} \;.
\label{Eq:SymmetricGauge-eta-constraint}
\end{align}
Using locality, each conjugate pair yields one linear equation for the $n$ unknowns $\eta_q$:
\begin{align}
\sum_{q=1}^n
\Big[
\chi_q(U_g P U_g^\dagger)-\chi_q(P)
\Big]\eta_q
=
\frac12
\Big(
x_{U_g P U_g^\dagger}^{\mathcal{G}_{\text{t}}}
-
x_P^{\mathcal{G}_{\text{t}}}
\Big) \;.
\label{Eq:SymmetricGauge-eta-linear}
\end{align}
Collecting sufficiently many independent conjugate-pair constraints gives a full-rank system for $\bm\eta$. Once $\bm\eta$ is known, the gauge-fixed SPAM parameters are obtained from Eqs.~\eqref{Eq:SymmetricGauge-gauge-s}--\eqref{Eq:SymmetricGauge-gauge-m}.

Because the symmetry results of the main text hold only approximately, Eq.~\eqref{Eq:SymmetricGauge-symmetric-condition} should be understood as a first-order gauge-fixing rule. In practice one may therefore solve Eq.~\eqref{Eq:SymmetricGauge-eta-linear} by least squares when multiple noisy symmetry constraints are available.

%% ============================================================
%% APPENDIX: Experimental details
%% ============================================================

\section{Experimental details: parallel CZ learning circuits}
\label{App:ExptDet}

This Appendix specifies the circuit construction, Pauli-tracking equations, and $X$ error-injection procedure used in the experiments reported in the main text.

%%%Fig: kingston layout %%%%%%%%%%%%%%%%%%%%%%%%%%%%
\begin{figure}[t!]
\centering
\includegraphics[scale=0.73]{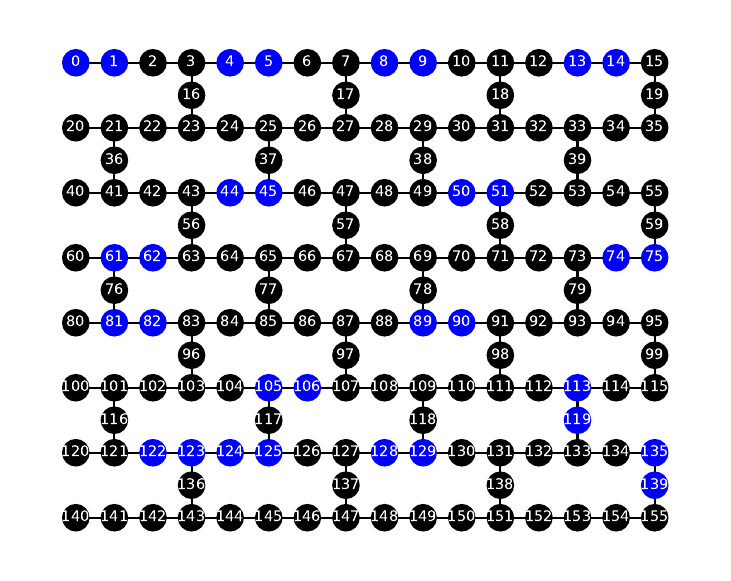}
\caption{\textbf{Layout of the CZ noise-learning experiment on IBM Kingston}---The sixteen non-overlapping qubit pairs selected from the device coupling map are highlighted in blue. The learning protocol in Sec.~\ref{SubApp:CircuitConstruct} is run on all pairs simultaneously.}    
\label{Fig:kingston_layout}
\end{figure}
%%%%%%%%%%%%%%%%%%%%%%%%%%%%%%%%%%%%%%%%%%%%%%%%%%%

\subsection{Circuit construction}
\label{SubApp:CircuitConstruct}

The learning protocol is implemented on the
IBM Kingston backend. Sixteen non-overlapping qubit pairs are selected from the device coupling map and executed simultaneously in a single circuit of width $2 \times 16 = 32$ qubits (Fig.~\ref{Fig:kingston_layout}).  Three circuit families are used:

\begin{itemize}
\item \textbf{Depth-0 (SPAM):} Prepare $\ket{0}^{\otimes 32}$ and measure in the computational basis.
\item \textbf{H0 (CZ):} Apply $H \otimes I$ on each pair, then CZ across all pairs simultaneously, then $H \otimes I$ again, followed by measurement.
\item \textbf{H1 (CZ):} Similar to H0, but with $I \otimes H$ in place of $H \otimes I$.
\end{itemize}

\noindent Hadamard conjugation rotates the effective measurement basis so that the depth-1 circuits probe Pauli fidelities outside the $Z$-diagonal sector.  All circuits are transpiled at optimization level~0 with a fixed initial layout mapping virtual qubits $(2i, 2i{+}1)$ to the physical pair~$i$.

\subsection{Pauli twirling and sampling}

Each transpiled circuit is Pauli-twirled (via \textsc{samplomatic}~\cite{qiskit_samplomatic}) by inserting random Pauli gates before and after each CZ layer, with corresponding classical bit-flips on the measurement outcomes. The \texttt{active\_circuit} strategy with the full two-qubit Pauli twirling group is used throughout.  For each base circuit, 500 independent randomizations are drawn and each is executed with 100 shots, giving $5 \times 10^4$ effective samples per
observable.

\subsection{Pauli tracking and equation formation}
\label{SubApp:ExperimentEquations}

For each qubit pair the observables of interest are $\{IZ, ZI, ZZ\}$.  We track each Pauli $P$ through the circuit using Clifford conjugation rules.

The depth-0 circuit yields
\begin{align}
y_{d0,P} = x^S_P + x^M_P \;, \qquad P \in \{IZ, ZI\},
\label{Eq:ExptDet-d0-eqs}
\end{align}
with $x^S_P \equiv -\log f^S_P$ and $x^M_P \equiv -\log f^M_P$.

For the H0 circuit, each observable $P$ is conjugated by $(I \otimes H) \to \mathrm{CZ} \to (I \otimes H)$.  Tracking through these layers, the Pauli that enters the CZ layer determines whether a gate fidelity appears. The resulting equations are

\begin{align}     
y_{h0,ZI} &= x^S_{ZZ} + x^{\mathcal{G}_{\text{t}}}_{XZ} + x^M_{ZI} \;, \nonumber\\
y_{h0,ZZ} &= x^S_{ZI} + x^{\mathcal{G}_{\text{t}}}_{XI} + x^M_{ZZ} \;.
\label{Eq:ExptDet-h0-eqs}
\end{align}

Similarly, the H1 circuit gives
\begin{align}
y_{h1,IZ} &= x^S_{ZZ} + x^{\mathcal{G}_{\text{t}}}_{ZX} + x^M_{IZ} \;, \nonumber\\
y_{h1,ZZ} &= x^S_{IZ} + x^{\mathcal{G}_{\text{t}}}_{IX} + x^M_{ZZ} \;.
\label{Eq:ExptDet-h1-eqs}
\end{align}

These six measurement equations~(\ref{Eq:ExptDet-d0-eqs})--(\ref{Eq:ExptDet-h1-eqs}) are supplemented by locality of the SPAM channels,
\begin{align}
x^S_{ZZ} &= x^S_{IZ} + x^S_{ZI} \;, \nonumber\\
x^M_{ZZ} &= x^M_{IZ} + x^M_{ZI} \;,
\label{Eq:ExptDet-locality}
\end{align}
and the symmetric-gauge conditions (Result~\ref{result:MultiPauliPhase2} applied to the CZ gate),
\begin{align}
x^{\mathcal{G}_{\text{t}}}_{XI} &= x^{\mathcal{G}_{\text{t}}}_{XZ} \;, \nonumber\\
x^{\mathcal{G}_{\text{t}}}_{IX} &= x^{\mathcal{G}_{\text{t}}}_{ZX} \;.
\label{Eq:ExptDet-symmetry}
\end{align}
The ten equations
(\ref{Eq:ExptDet-d0-eqs})--(\ref{Eq:ExptDet-symmetry})
determine the ten unknowns
($x^S_{IZ}$, $x^S_{ZI}$, $x^S_{ZZ}$, $x^M_{IZ}$, $x^M_{ZI}$,
$x^M_{ZZ}$, $x^{\mathcal{G}_{\text{t}}}_{XI}$, $x^{\mathcal{G}_{\text{t}}}_{XZ}$,
$x^{\mathcal{G}_{\text{t}}}_{IX}$, $x^{\mathcal{G}_{\text{t}}}_{ZX}$) and the system
admits a unique closed-form solution.

\subsection{Post-processing}

Each job returns bitstrings of length 32 (two classical bits per pair).  For pair~$i$ we extract the two-bit marginal at positions $(2i, 2i{+}1)$.  Readout-twirling flips are applied to the raw bitstrings before computing expectation values.  For a
$Z$-type observable with binary mask $\bm{z}$, the
single-randomization estimator is
\begin{align}
\langle P \rangle
= \frac{1}{N_{\mathrm{shots}}}
  \sum_{\text{shots}} (-1)^{\bm{z} \cdot \bm{b}} \;,
\end{align}
where $\bm{b}$ is the corrected bitstring after readout twirling.  The mean and standard error are computed over the 500 randomizations. Log-expectation values $y = -\log\langle P \rangle$ are then substituted into the closed-form solution, with uncertainties propagated analytically.

\subsection{Synthetic error-injection protocol}

Controlled incoherent state-preparation errors are constructed by
mixing experimental data collected under two preparation
settings for each qubit pair: the identity $I$, and a bit-flip on
the first qubit $X$.
The noise injection is implemented by first Pauli twirling and then inserting an $X$ gate at the beginning of the circuit, directly on the physical qubit indices of the transpiled circuit, so it is not absorbed by the twirling layer and acts as a genuine state-preparation error.  For each setting all three circuit families (depth-0 and two depth-1) are executed with the same twirling randomizations, producing $3 \times 2 = 6$ primitive unified blocs (PUBs) submitted in a single job.

The synthetic bit-flip channel at strength~$p$ is then
constructed in post-processing by mixing the measured expectation values as in Eq.~\eqref{eq:synth_mixing}.  A one-dimensional sweep of the synthetic bit-flip probability 
$p$ is then performed at 21 equally spaced values $p \in [0, 0.3]$.  The full SPAM-gate decomposition is carried out independently for each (pair, $p$) combination and for each gauge choice.

\subsection{Physicality self-verification}
All Pauli fidelities extracted by the gauge decomposition must
satisfy $f \le 1$; a violation places the solution outside the
physical domain and signals that statistical noise, calibration
drift, or a breakdown of the model assumptions (e.g.\
non-Markovianity or violation of the symmetric-gauge condition)
has corrupted the estimate for that qubit pair.

We flag a pair as unreliable whenever \emph{any} of its extracted fidelities, i.e., state preparation, measurement, or
gate, exceeds unity at $p = 0$.  In Fig.~\ref{fig:spam_errors}
such pairs are marked by red vertical bands spanning both qubits. Pairs that pass the physicality check exhibit clean SPAM-gate separation across the full $p$~sweep, as illustrated in Fig.~\ref{fig:synth_validation}.

\section{Comparison with prior SPAM characterization methods}
\label{app:comparison}

Yu and Wei~\cite{yu2025efficient} proposed two approaches for separately quantifying state-preparation (SP) and measurement errors using an ancilla qubit ($q_a$) coupled to the qubit of interest ($q_t$) via a noisy CNOT gate.  We briefly summarize both methods, then present a numerical comparison with the
gauge-optimization approach introduced in the main text.

\subsection{Symmetrization method}

The main idea of this approach is that using a CNOT gate, one can measure additional combination of state preparation and measurement errors that would enable resolving them individually. We first follow the notation in Ref.~\cite{yu2025efficient} to introduce the method and then connect it to our formalism. 

%%%%%%%%%%% Fig: Comparison to prior work %%%%%%%%%%%%%%%%%%%%%%
\begin{figure*}[t!]
  \centering
  \includegraphics[width=\textwidth]{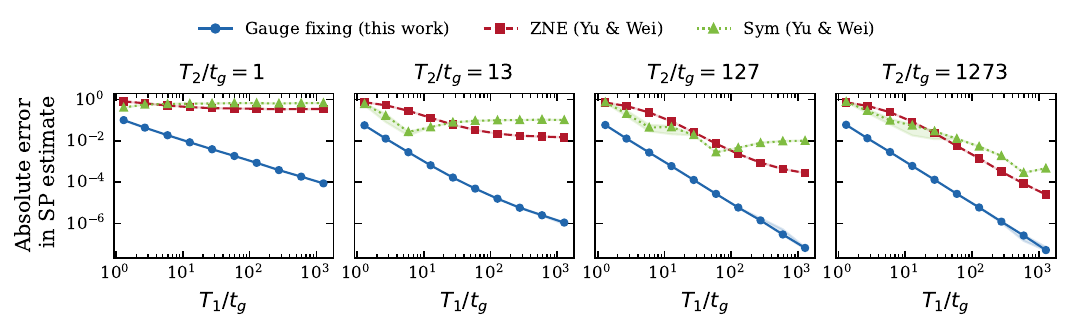}
  \caption{Absolute error in the inferred state-preparation parameter as a function of $T_1/t_g$ for four values of $T_{2\phi}/t_g$. Lines show the median over 10 random SPAM realizations per point; shaded bands span the interquartile range. The symmetrization method (green, dotted) develops a large systematic bias whenever the gate noise is appreciable ($T_1/t_g \lesssim 100$), because the Hadamard conjugation used to reverse the native~CX breaks the noise-symmetry assumption. Zero-noise extrapolation (red, dashed) removes most of this bias but its linear extrapolation becomes inaccurate at large error rates. The gauge method (blue, solid) remains accurate across the full parameter range.}
  \label{fig:comparison}
\end{figure*}
%%%%%%%%%%%%%%%%%%%%%%%%%%%%%%%%%%

A single-qubit readout is described by a two-element POVM with bit-flip
probabilities~$\delta_M^0$ (reading~$|0\rangle$ as~$|1\rangle$) and
$\delta_M^1$ (reading~$|1\rangle$ as~$|0\rangle$), and a faulty
preparation of~$|0\rangle$ produces the incoherent state
$\rho = (1-\delta_\mathrm{SP})\,|0\rangle\!\langle 0|
      + \delta_\mathrm{SP}\,|1\rangle\!\langle 1|$.
The SPAM errors are~\cite{yu2025efficient}
\begin{align}
  \delta_\mathrm{SPAM}^{0}
    &= (1 - \delta_M^0 - \delta_M^1)\,\delta_\mathrm{SP}
       + \delta_M^0\,,
  \label{eq:spam0} \\
  \delta_\mathrm{SPAM}^{1}
    &= (1 - \delta_M^0 - \delta_M^1)\,\delta_\mathrm{SP}
       + \delta_M^1\,.
  \label{eq:spam1}
\end{align}

Now consider a noiseless CNOT with~$q_t$ as the control and~$q_a$ as
the target, both initialized in~$|0\rangle$.  With
probability~$\delta_\mathrm{SP}^t$ the control qubit is in~$|1\rangle$,
which flips the ancilla via the CNOT.  
Let $\delta_\mathrm{SP}^a$ denote the bare preparation error of the
ancilla. Ideally, the ancilla qubit should be measured in $\ket{0}$. However, state
preparation errors may leave it in the wrong state. The resulting effective
SPAM error on the ancilla qubit is then~\cite{yu2025efficient}
\begin{equation}
  \tilde\delta_\mathrm{SPAM}^{a,0}
    = (1 - \delta_\mathrm{SPAM}^{a,0}
         - \delta_\mathrm{SPAM}^{a,1})\,\delta_\mathrm{SP}^t
      + \delta_\mathrm{SPAM}^{a,0}\,.
  \label{eq:spam_cx}
\end{equation}

We now specialize the above result from Ref.~\cite{yu2025efficient} to the case with symmetric readout errors
($\delta_M^0 = \delta_M^1 \equiv \delta_M$),
Eqs.~\eqref{eq:spam0}--\eqref{eq:spam1} give
$\delta_\mathrm{SPAM}^{0} = \delta_\mathrm{SPAM}^{1}$, and solving
Eq.~\eqref{eq:spam_cx} for~$\delta_\mathrm{SP}^t$ yields
\begin{equation}
  \delta^t_\mathrm{SP}
  = \frac{\tilde\delta_\mathrm{SPAM}^{a,0}
          - \delta_\mathrm{SPAM}^{0}}
         {1 - 2\,\delta_\mathrm{SPAM}^{0}}\,.
  \label{eq:sym}
\end{equation}
In the presence of a noisy CNOT gate, Ref.~\cite{yu2025efficient} shows that
Eq.~\eqref{eq:sym} continues to hold, provided the gate noise
acts symmetrically on the control and target qubits.

In practice, however, current superconducting processors provide a
native CNOT in only one direction.  The reversed CNOT is obtained by
conjugating with Hadamard gates on both qubits,
$\mathrm{CX}_{01} = (H\!\otimes\! H)\,\mathrm{CX}_{10}\,(H\!\otimes\! H)$.
Because the physical gate noise~$\mathcal G$ originates from the
fixed native interaction, the reversed circuit sees the
conjugated noise channel
$(H\!\otimes\! H)\,\mathcal{G}\,(H\!\otimes\! H)$ rather
than~$\mathcal G$ itself. As it was also observed in Ref.~\cite{yu2025efficient}, the assumption about the direction of noise therefore reduces to an assumption about the symmetry between the $Z$ and $X$ components of the noise channel. To see this clearly, we simplify Eq.~\eqref{eq:sym} and express in terms of the Pauli fidelities of the noise channel. We then find that 
\begin{equation}
    % \label{eq:sym-fidpicture}
    \delta^t_\mathrm{SP}=\frac{1}{2}\left(1- \frac{f^{\mathcal{G}}_{ZI}}{f^{\mathcal{G}}_{XI}}\frac{f^{S}_{ZZ}}{f^S_{ZI}}\right).
\end{equation}
Noting that with the local SPAM assumption we have $f^{S}_{ZZ}=f^{S}_{IZ}f^{S}_{ZI}$, we find that
\begin{equation}
    \label{eq:sym-fidpicture}
    \delta^t_\mathrm{SP}=\frac{1}{2}\left(1- \frac{f^{\mathcal{G}}_{ZI}}{f^{\mathcal{G}}_{XI}}f^S_{IZ}\right).
\end{equation}
Since $2\delta^t_\mathrm{SP}=1-f^S_{IZ}$, this relationship clearly shows that the protocol of Ref.~\cite{yu2025efficient} requires $f^{\mathcal{G}}_{ZI}=f^{\mathcal{G}}_{XI}$. However, for common error sources such as amplitude damping and dephasing, even when the error rates on the qubits are identical, these two fidelities do not match. Therefore, as we also demonstrate numerically in the following, this assumption typically introduces a non-negligible bias in the state-preparation error estimate.  

\subsection{Zero-noise extrapolation
  (Appendix~B of Ref.~\cite{yu2025efficient})}

To mitigate the bias from gate noise, Ref.~\cite{yu2025efficient} proposes
repeating the noisy CNOT gate an odd number of times
$m = 1,3,5,\dots$ (so that the net unitary remains a single~CX) and
measuring the ancilla SPAM error at each~$m$.  The gate-noise
contribution grows with~$m$, while the SP error contribution is
independent of~$m$.  A linear extrapolation to $m = 0$ removes the
gate-noise bias and isolates $\delta_\mathrm{SP}$.

This approach does not explicitly make an assumption about gate-noise symmetry but
requires additional circuits for each value of~$m$ and relies on the
noise scaling linearly with the number of gate repetitions, an
approximation that degrades at large error rates.

\subsection{Numerical comparison}

We compare the two methods above with the gauge-optimization approach
described in the main text on a Lindblad simulation of a noisy CX gate.
The simulation proceeds as follows.

The two-qubit gate is generated by the Hamiltonian
\begin{equation}
  H_\mathrm{CX}
  = XI - XZ - II + IZ\,,
\end{equation}
which produces $\mathrm{CX}_{10}$ (control on the second qubit, target on
the first) via $U = e^{-i(\pi/4) H_\mathrm{CX}}$. Amplitude damping ($T_1$) and pure dephasing ($T_{2\phi}$) are modeled as
Lindblad dissipators with dissipators $\mathcal{D}[S^{-}]$ and $\mathcal{D}[Z]$ at rates $\gamma_1$ and $\gamma_{2\phi}$, respectively, applied identically to both qubits.  The noisy gate channel is obtained by exponentiating the
full Lindbladian for one gate time ($t_g = \pi/4$) and removing the
ideal unitary, yielding a diagonal Pauli channel with fidelities
$\{f_P\}$.

State-preparation and measurement errors are modeled as independent
single-qubit depolarizing channels.  For each trial, the depolarizing
parameters of both the SP and measurement channels are drawn uniformly
from $[0, 0.1]$ on each qubit.  All three methods receive the same
Pauli-diagonal channel data and attempt to recover the SP depolarizing
parameter of the control (second) qubit.

We sweep the decay rates over a logarithmic grid with
$T_1/t_g, T_{2\phi}/t_g \in \{1.3, 2.7, \dots, 1273\}$ (10~points each) and average over 10~random SPAM realizations per grid point.  Fig.~\ref{fig:comparison}
shows the median absolute error of each method as a function of $T_1/t_g$
at four representative values of $T_{2\phi}/t_g$, with shaded bands indicating the interquartile range.

The gauge-fixing method achieves the lowest error across the entire parameter space.  The symmetrization method incurs a systematic bias that grows as
$T_1/t_g$ decreases (i.e.\ as the gate noise increases), consistent
with the violation of the noise-symmetry assumption discussed above. The ZNE method substantially reduces this bias but still degrades at strong noise, where the linear extrapolation in the number of gate
repetitions becomes inaccurate.  

\bibliography{gauge_spam_bibliography}
\end{document}